\documentclass[%
aip,
amsmath,amssymb,
reprint,%
]{revtex4-1}
\usepackage{graphicx}
\usepackage{dcolumn}
\usepackage{bm}
\usepackage{amsmath}
\usepackage{xspace}
\usepackage{listings}
\usepackage{xcolor}
\usepackage[utf8]{inputenc}
\usepackage[T1]{fontenc}
\usepackage{mathptmx}
\usepackage{listings}
\usepackage{adjustbox}
\usepackage[normalem]{ulem}
\definecolor{codegreen}{rgb}{0,0.4,0}
\definecolor{codeblue}{rgb}{0.,0.,0.7}
\definecolor{codegray}{rgb}{0.5,0.5,0.5}
\definecolor{codepurple}{rgb}{0.58,0,0.82}
\definecolor{backcolour}{RGB}{248,248,250}
\lstdefinestyle{mystyle}{
  language=Python,
  commentstyle=\color{codeblue},
  numberstyle=\tiny\color{codegray},
  stringstyle=\color{codegreen},
  basicstyle=\linespread{1.}\ttfamily\small,
  breakatwhitespace=true,         
  breaklines=true,                 
  captionpos=b,                    
  keepspaces=true,                 
  numbers=none,                    
  numbersep=5pt,                  
  showspaces=false,                
  showstringspaces=false,
  showtabs=false,                
  tabsize=4
}

\usepackage{comment}
\usepackage{multirow,booktabs,caption,threeparttable}
\usepackage{color}

\begin{document}
\title{
Enhancing PySCF-based Quantum Chemistry Simulations with Modern Hardware,
Algorithms, and Python Tools
}
\author{Zhichen Pu}
  \altaffiliation{Bytedance Seed}
\author{Qiming Sun}
  \email{qiming.sun@bytedance.com.}
  \altaffiliation{Bytedance Seed}

\begin{abstract}
\noindent
The PySCF package has emerged as a powerful and flexible open-source platform
for quantum chemistry simulations. However, the efficiency of electronic
structure calculations can vary significantly depending on the choice of
computational techniques and hardware utilization. In this paper, we explore
strategies to enhance research productivity and computational performance in
PySCF-based simulations. First, we discuss GPU acceleration for selected PySCF
modules. Second, we demonstrate algorithmic optimizations for particular
computational tasks, such as the initial guess manipulation, the second-order
self-consistent field (SOSCF) methods,
multigrid integration, and density fitting approximation, to improve convergence
rates and computational efficiency. Finally, we explore the use of modern Python
tools, including just-in-time (JIT) compilation and automatic differentiation to
accelerate code development and execution.
These approaches present a practical guide for enhancing the use of PySCF's
capabilities in quantum chemistry research.
\end{abstract}
\maketitle

\section{Introduction}
\label{sec:intro}
Quantum chemistry simulations play an important role in modern chemistry research.
Advancements in software engineering and computing hardware have led to the emergence of a new generation of quantum chemistry computation packages and tools,
offering more capabilities in quantum chemistry simulations.
Some packages prioritize ease of use by employing user-friendly programming
languages like Python, which enhances code readability\cite{Sun2020,Smith2018,Boguslawski2021,Hermann2016,Gao2025,Evangelista2024,Chan2024,Posenitskiy2023}.
Some tools leverage advancements in computing models, such as automatic
differentiation and tensor contractions, to improve the efficiency of
developing new methods\cite{Zhang2022,Kasim2022,Matteo2023,Zhou2020,Abbott2021,Friede2024}.
Some are optimized for specific hardware, like GPUs, to maximize
performance and handle more complex calculations\cite{Wu2025,Kim2024,Tornai2019,Li2025a,Vallejo2023,Seritan2021}.
There are also ongoing efforts to experiment with modern programming languages such as Julia and Rust to incorporate the latest software
innovations\cite{Poole2020,Aroeira2022,Wang2023,Wang2024}.

Balancing performance and simplicity is a challenge in the development of quantum chemistry simulation program.
Historically, quantum chemistry packages heavily prioritized performance, often at the expense of code readability and engineering design.
Developing such quantum chemistry programs has a high barrier as it requires dedicated efforts and programming skills.
For the time being, code readability and program maintainability receive more attention, as the functionalities and methodologies in quantum chemistry become more complicated.
With the evolution of programming languages and software engineering concepts, the adoption of object-oriented programming languages like C++ makes programs more modular and manageable.
Several quantum chemistry packages, such as ORCA\cite{Neese2020}, MPQC\cite{Peng2020}, Psi4\cite{Smith2020},
Bagel\cite{Shiozaki2018}, and Chronous\cite{Williams-Young2020}, have been written in C++, which has significantly enhanced code readability and maintenance.
This shift has made quantum chemistry software more accessible and more suitable for developing complex methods.

Quantum chemistry simulations were used to be treated as black boxes, which is suitable for simple and straightforward computational tasks.
However, as simulation tasks become more complex and demanding,
especially when quantum chemistry research integrates with machine learning (ML) techniques
\cite{Dral2020,Gong2025,Chen2024,Khan2025,Wu2023,Kirkpatrick2021,Li2021,Kasim2021}, tasks such as adjusting programs for hybrid theoretical models, managing computation workflows, and developing effective data processing programs are becoming inevitable.
Despite the advancements, the rigidity of C++ quantum chemistry packages often
requires users to acquire substantial C++ programming skills, which can pose a significant barrier to entry.
In contrast, the Python programming language is renowned for its ease of understanding and learning, making Python programs more straightforward to comprehend and extend.
Additionally, due to its popularity in artificial intelligence (AI) development, many attractive tools have been developed by the AI community, granting Python exceptional usability for building scientific computation tools.
An increasing number of quantum chemistry tools are adopting Python, and many challenging research projects are being developed and solved using Python code.
These trends reflect a shift towards more accessible and flexible programming environments.

PySCF was originally designed to serve this role, offering flexibility for testing new ideas and challenges.
To balance simplicity and performance, PySCF was implemented primarily in Python,
with performance-critical hotspots optimized in the C language\cite{Sun2018,Sun2020}.
This is a common design pattern in high-performance Python packages\cite{Sun2020,Evangelista2024,Smith2020}.
This technology stack enables PySCF to maintain ease of understanding, while still retaining the computational performance for complex quantum chemistry simulations.

When applying PySCF in more chemistry simulation applications, some doubts were
raised by users about its computation efficiency and capabilities.
Due to the optimization strategies and the API (application program interface) design of the package, the default methods provided by PySCF may not always present the optimal performance.
There are various techniques to leverage PySCF's built-in modules to enhance performance.
However, these techniques are not extensively documented in PySCF official documentation, examples, or published papers.
Addressing this gap is one of the motivations for writing this paper, which aims to provide a guide for accelerating computations using the functionalities provided by PySCF.
This includes the use of GPU acceleration, improved initial guesses, effective self-consistent-field (SCF) convergence strategies, and efficient integral computation routines available in PySCF.

Moreover, PySCF is designed as a Python package rather than standalone software.
This design allows PySCF to be seamlessly integrated into any existing Python infrastructure.
PySCF can be used with various Python tools to enhance its capabilities.
The functionalities provided by the Python community offer broader ways to improve the productivity of quantum chemistry simulations.
This is another point we aim to illustrate in this work.

In this paper, we will discuss several techniques to enhance the productivity of the PySCF package.
Section~\ref{sec:overview} briefly reviews the functionalities of PySCF package and its ecosystem.
Section~\ref{sec:gpu4pyscf} demonstrates how to accelerate computations using the PySCF GPU extension.
Section~\ref{sec:pyscf:algorithms} discusses how to improve computation
efficiency using functionalities provided by the PySCF package.
Section~\ref{sec:python:tools} explores how PySCF can be combined with
just-in-time compilation and automatic differentiation techniques.

\section{Overview of PySCF}
\label{sec:overview}
\subsection{Features provided by PySCF}
PySCF is an open-source Python package for solving electronic structure problems
using Gaussian-type basis sets.
It supports both molecular systems and extended systems with periodic boundary conditions.
Key features and functionalities of PySCF include:
\begin{itemize}
  \item Mean-field energy, gradients, and Hessian.
  \item Single-reference post-Hartree-Fock methods such as M{\o}ller-Plesset
    perturbation, coupled cluster,
    and Random phase approximation (RPA) methods.
  \item Multi-configurational and multi-reference methods for electron correlation.
  \item Excited states modelled by density functional theory,
    equation-of-motion coupled cluster, and ADC (Algebraic Diagrammatic Construction) methods.
  \item Various molecular properties calculations.
  \item Scalar relativistic, two-component and four-component relativistic effects.
  \item Solvent effects through implicit solvent models.
  \item DFT and post-HF methods for extended systems with periodic boundary condition.
\end{itemize}

\subsection{PySCF Eco-system}
The PySCF package is organized into core modules, PySCF-Forge, and various extensions.
The core module provides the fundamental functionalities required for quantum chemistry simulations.
Within this module, settings and strategies are implemented conservatively, with tight thresholds to ensure accurate and reliable results.
The implementation of the PySCF core module focuses on functional programming, lightweight design yet efficient performance.

Newly developed methods are available in PySCF-Forge, which serves as a staging
area for future features in PySCF.
GPU acceleration is implemented in the GPU4PySCF package\cite{Li2025a,Wu2025}, which is specifically
designed to enhance performance.
The use of GPU4PySCF with PySCF will be discussed in Section~\ref{sec:gpu4pyscf}.
Another significant extension is PySCFAD\cite{Zhang2022}, which extends PySCF by
providing automatic differentiation capabilities for all methods in PySCF.
These are the primary extensions of PySCF.
Additional extensions, which may be written in different languages or managed
at different levels of maintenance, are available on the PySCF organization page on GitHub.
We will not list all of them in this paper.
The core module, PySCF-forge and these extensions constitute the PySCF ecosystem.

\subsection{PySCF APIs}
PySCF is designed as a Python package.
Unlike traditional quantum chemistry packages, which typically use a DSL
(domain specific language) configuration file through a driver,
PySCF provides APIs to execute its functionalities directly in Python.

Quantum chemistry methods are implemented as classes, organized in the object-oriented programming (OOP) structure.
To perform a quantum chemistry task, one can instantiate a \verb$Mole$ object,
apply the classes corresponding to the desired methods, and then call the
\verb$.run()$ method to execute the computation. For example
\begin{verbatim}
mol = pyscf.M(atom='''
O    0.    0.000    0.118
H    0.    0.755   -0.471
H    0.   -0.755   -0.471''',
basis='cc-pvdz')
mol.RKS(xc='pbe0').run()
\end{verbatim}
Subsequent methods can be chained together to streamline computations.
For instance, the time-dependent density functional theory (TDDFT) methods and
their gradients can be executed sequentially:
\begin{verbatim}
mf = mol.RKS(xc='pbe0').run()
mf.TDDFT().run().Gradients().run()
\end{verbatim}
Parameters for each method can be specified as keyword arguments when calling the \verb$.run()$ method,
or be assigned directly to each instance, just like manipulating the attributes of a class in OOP programming.
For example, the following two setups are equivalent:
\begin{widetext}
\begin{lstlisting}[style=mystyle]
# Supply parameters as keyword arguments
mol.RKS(xc='pbe0').run().TDDFT().run(nstates=6, triplet=False).Gradients().run()
# Assigin paramteres to attributes
mf = mol.RKS()
mf.xc = 'pbe0'
mf.run()
td = mf.TDDFT()
td.nstates = 6
td.triplet = False
td.run()
td_grad = td.Gradients()
td_grad.run()
\end{lstlisting}
\end{widetext}
Additional features, such as solvent model, relativistic
corrections, can be applied within the chain call as well.
For example, the PCM solvent model are attached to the DFT calculation
and subsequently applied to the TDDFT excited state computation:
\begin{verbatim}
mol.RKS(xc='pbe0').PCM().run().TDDFT().run()
\end{verbatim}
The solvent model is applied to the RKS method.
Its effects influence the ground state wavefunction (the DFT orbitals), as
well as the subsequent TDDFT response calculation.
If the application order is altered,
\begin{verbatim}
mol.RKS(xc='pbe0').run().TDDFT().PCM().run()
\end{verbatim}
the orbitals from gas phase will be used in the linear response computation.
Although the solvent model will still contribute to the linear response matrix,
the results will be different from those obtained using the RKS-PCM-TDDFT combination.
Unlike writing configuration in the input file where feature keywords are simply listed,
the order in which methods are applied matters using the chain execution APIs.

The second type of APIs in PySCF are individual functions.
They provide independent features that do not belong to a specific simulation method.
A typical example of this category is the integral evaluator function.
The common API \verb$mol.intor()$ method supports the evaluation of various types
of integrals.
The \verb$pyscf.scf.jk.get_jk()$ function is another example, which provides the general
function to compute Coulomb and exchange matrices.
It supports an arbitrary number of density matrices and any two-electron integrals
implemented by the Libcint library\cite{Sun2015}, including zeroth-order,
high-order derivatives, or Gauge-included two-electron integrals, across multiple molecules.
For more details on these functions, please refer to the PySCF online
documentation at \url{www.pyscf.org}.


\section{GPU Acceleration}
\label{sec:gpu4pyscf}
To maintain functional generality and ease of modification, the default
algorithms and engineering practices in PySCF may not be optimized for peak performance.
Additionally, to decrease numerical uncertainty, various thresholds in PySCF are set very tight.
These choices lead to slower execution for certain functionalities when using the default code in PySCF, compared to other packages.
To address this problem, one method is to select different algorithms than the default one, which will be discussed in Section~\ref{sec:pyscf:algorithms}.
Another option is to switch to the more efficient package, GPU4PySCF, to achieve
faster execution.

The GPU4PySCF extension is developed with a priority on performance optimization.
As the name indicates, the package leverages GPU hardware to improve the processing of computationally expensive workloads.
This includes the implementation of various integral computation algorithms on GPU and the use of GPU tensor libraries for linear algebra operations.
When designing algorithms and their implementations, this extension prioritizes performance-oriented variants.
The code readability and the generality of features may not be as emphasized as they are in PySCF core modules.

At the current stage, the GPU4PySCF package primarily focuses on accelerating DFT computations.
The most significant performance improvements are observed in two areas.
The first is DFT calculations for medium-sized molecules (50 - 100 atoms).
The second area is in handling large molecules that involve over 10,000 basis functions.
For medium-sized molecules, the density fitting approximation is employed.
Thanks to the substantial advantages of tensor contraction operations on GPUs,
GPU4PySCF can achieve speedups of 2 - 3 orders of magnitude over traditional CPU-based code\cite{Wu2025}.

When it comes to molecules that contain over 1,000 atoms, or more than 10,000
basis functions, performing standard DFT calculations becomes a significant
challenge for many CPU-based programs.
Even if computationally feasible, the time required to complete these
calculations is impractical for routine tasks.
In the case of GPU programs, although GPUs offer substantial computational capabilities, there are technical challenges in performing these large-scale calculations on GPUs.
The density fitting approximation is not feasible due to its high scaling when evaluating the HF exchange matrix.
Integrals must be evaluated on the fly during the SCF iteration.
In this context, various analytical integral algorithms have been intensively optimized within the GPU4PySCF package for the calculation of Coulomb and exchange matrices\cite{Li2025a}.
The technical details of the program implementation are beyond the scope of this paper.
More details will be presented in our upcoming works.

The GPU4PySCF APIs generally follows the APIs provided by the PySCF package,
including the naming conventions, the order of parameters, the keyword arguments in function signatures.
GPU4PySCF also implements Python classes to represent the individual quantum chemistry methods
using the same name as those in PySCF core, as well as the parameters and attributes within each method class.
This design allows users to seamlessly replace the PySCF core implementations with those of GPU4PySCF in a straightforward manner.
For instance, the restricted Kohn-Sham DFT method in the PySCF core module is
available at \verb$pyscf.dft.RKS()$.
In GPU4PySCF, this method is accessible as
\verb$gpu4pyscf.dft.RKS()$.
To perform a RKS calculation using GPU4PySCF, the script can be written
following the PySCF APIs:
\begin{verbatim}
from gpu4pyscf.dft import RKS
RKS(mol, xc='pbe0').run()
\end{verbatim}

To simplify the conversion between GPU4PySCF and PySCF objects,
two methods, \verb$.to_gpu()$ and \verb$.to_cpu()$, have been introduced in PySCF and GPU4PySCF classes respectively.
For an instance created by PySCF code, the \verb$.to_gpu()$ method
recursively transforms the PySCF instance and its attributes to objects suitable for GPU acceleration.
After executing the computation on GPU, the object can be transformed back to
the instance of the PySCF core modules, simplifying the subsequent CPU operations.
For example, we can execute the computationally intensive TDDFT diagonalization
on GPU and then analyze the TDDFT response coefficients using the
functionalities provided by the PySCF core on CPU.
  \begin{verbatim}
tddft_on_gpu = mol.RKS().to_gpu().run().TDA()
tddft_on_gpu.to_cpu().analyze()
\end{verbatim}

In addition to the classes for individual simulation models, GPU4PySCF also offers
several functions to accelerate the demanding functions implemented in PySCF, particularly those for integral computation.
For example, the \verb$get_jk$ function in the \verb$gpu4pyscf.scf.hf$
module can efficiently evaluate the Coulomb matrix and exchange matrices for any given
density matrices.
\begin{verbatim}
from gpu4pyscf.scf.hf import get_jk
jmat, kmat = get_jk(mol, dm)
\end{verbatim}
This function can work as a drop-in replacement to the \verb$get_jk$ function of the \verb$pyscf.scf.hf$ module.
Another example is the numerical integration functions provided by the \verb$gpu4pyscf.dft.numint$ module.
In this module, density evaluation and integral computations functions for
various types of exchange-correlation (XC) functionals have been implemented using
custom GPU kernels and tensor contraction code on GPU.
The numerical integration features in PySCF can also be enhanced using these accelerated functions.
\begin{verbatim}
from gpu4pyscf.dft.numint import NumInt
rho = NumInt().get_rho(mol, dm, grids)
\end{verbatim}
Other function-level acceleration features include the evaluation of three-center integrals and their derivatives,
grid-based Coulomb integral evaluation in the implicit solvent model,
the calculation of the Coulomb interaction matrix between the quantum mechanical
(QM) region and molecular mechanics (MM) particles in the QM/MM model\cite{Li2025},
and the multigrid algorithm for DFT XC integrals.
These functions can also be utilized to accelerate specific functions within the corresponding core modules of PySCF.

To minimize the overhead associated with data transfer between CPU and GPU
architectures, GPU4PySCF functions tend to store large-sized arrays, such as
wave-function coefficients, density matrices, and integrals, in CuPy arrays.
Using NumPy code to manipulate these CuPy arrays can lead to compatibility issues.
Therefore, when PySCF core functions are used to handle data generated by GPU4PySCF,
it is important to explicitly convert the data types of GPU4PySCF outputs using the CuPy function \verb$cupy.asnumpy()$.

\section{Methods and Algorithms to Improve PySCF Performance}
\label{sec:pyscf:algorithms}
\subsection{Manipulating Initial Guess}
The initial guess is crucial in electronic structure simulations.
The quality of the initial guess can significantly impact the computational performance and the correctness of the results.

The method classes in the PySCF package generate the default initial guess based on established experience.
It also allows users to construct their own initial guesses and pass them to each computational method.
In PySCF, custom initial guesses for mean-field methods, post-Hartree-Fock
methods, and excited state methods are handled in three separate approaches.

\subsubsection{Initial Guess for Mean-field Methods}
\label{sec:mf:init:guess}
For mean-field methods, the initial guess can be provided as a density matrix through the \verb$.kernel()$ method:
\begin{verbatim}
mf = mol.RKS(xc='pbe0')
mf.kernel(dm0=initial_guess_density_matrix)
\end{verbatim}
The density matrix for an initial guess does not necessarily originate from a single determinant.
There is no need to consider whether the density matrix obey the idempotent property or yield the correct number of integer electrons.
It can be constructed simply to reflect specific chemical intuition or the locality of the chemical system.

\begin{figure}
  \begin{center}
    \includegraphics[width=0.3\textwidth]{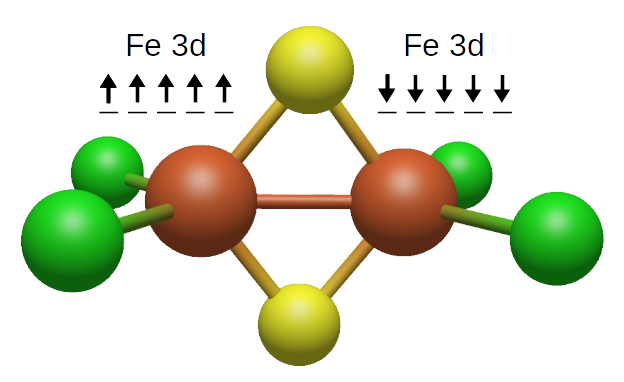}
  \end{center}
  \caption{The electron spin are customized in the initial guess for the Fe atoms in the Fe$_2$S$_2$Cl$_4$ molecule.}
  \label{fig:iron:sulfur}
\end{figure}

To customize the initial guess density matrix, it is necessary to know how the
basis functions are ordered, as different software packages adopt different conventions.
To eliminate any ambiguity in the order of basis functions, the PySCF package
offers several APIs to help users identify the atomic character of the basis functions.
The \verb$.ao_labels()$ method of the \verb$Mole$ class provides the labels of
each basis function, such as \verb$'1 C 4py'$.
The three components in the output represent the atom ID, the symbol of the element, and the basis notation, respectively.
The \verb$.search_ao_label()$ method of the \verb$Mole$ class returns the
indices of the basis functions that match the specified regular-expression
pattern.
For example, \verb$mol.search_ao_label('C.*pz')$
would search for all $p_z$ type atomic orbitals (AO) associated with Carbon atoms within the molecule.
The following example demonstrates how to use these functions to construct a
spin-symmetry broken initial guess for the two Fe atoms in an iron-sulfur
molecule, as shown in Figure~\ref{fig:iron:sulfur}.
\begin{verbatim}
mf = mol.UKS(xc='pbe')
dma, dmb = mf.get_init_guess()
fe0_3d = mol.search_ao_label('0 Fe 3d')
fe1_3d = mol.search_ao_label('1 Fe 3d')
dma[fe0_3d[:,None],fe0_3d] = np.eye(5)
dmb[fe1_3d[:,None],fe1_3d] = np.eye(5)
dma[fe1_3d[:,None],fe1_3d] = 0
dmb[fe0_3d[:,None],fe0_3d] = 0
mf.kernel(dm0=np.array((dma, dmb)))
\end{verbatim}

Another effective practice for setting up the initial guess in mean-field
calculations is to project the initial guess from the results of a
smaller-scale, preliminary calculation.
This approach is particularly useful in potential energy surface scans and geometry optimization.
In these scenarios, one simply constructs the density matrix from a previous
calculation using the \verb$.make_rdm1()$ method, and then passes it to the next calculation.

When handling challenging systems such as gapless systems, ions with
diffused basis functions, or transition metal complexes with various spin-polarized
local minima, convergence issues may arise in regular SCF iterations.
In these scenarios, we can first converge the calculation using a small basis set (such as the DZ basis set without diffused functions) and a different functional (such as UHF).
The results can then be projected to larger basis sets using the \verb$project_dm_nr2nr()$ function from the \verb$pyscf.scf.addons$ module. For example,
\begin{verbatim}
from pyscf.scf.addons import project_dm_nr2nr
dma_dz, dmb_dz = mol_dz.UHF().run().make_rdm1()
dma_tz = project_dm_nr2nr(mol_dz, dma_dz, mol_tz)
dmb_tz = project_dm_nr2nr(mol_dz, dmb_dz, mol_tz)
mf = mol_tz.UKS(xc='pbe')
mf.kernel(dm0=(dma, dmb))
\end{verbatim}

If both the basis set and the geometry of the target system differ from those
used in the preliminary calculation, the projection should be performed between
the different basis sets while keeping the geometry unchanged.
The resulting density matrix can then be directly utilized in the full-sized problem.
Projecting the density matrix between molecules with different geometries should be avoided.
Doing so would result in the density matrix on the new geometry representing the electron density of the old geometry, which would be an inappropriate initial guess.
This should be avoided.

\subsubsection{Initial Guess for Post-Hartree-Fock Methods}
The initial guess for post-Hartree-Fock methods, particularly the configuration interaction (CI) family, requires a different format of initial guess.
Typically, the CI coefficients are derived from the MP2 amplitudes, which are generally sufficient for initializing both CI and coupled cluster (CC) methods.

Preparing the initial guess for multi-configurational self-consistent field (MCSCF) calculations is a special case.
It requires to supply both the orbitals and the wavefunctions of the active space solver (by default, the FCI solver).
In most scenarios, supplying an initial guess for the active space solver is not necessary, as it has relatively little impact on the results of the MCSCF problem.
In contrast, the initial orbitals play a crucial role in MCSCF calculations.
A proper set of initial orbitals not only accelerates convergence but also influences the correctness of the results.
Various methods have been explored for constructing MCSCF initial orbitals\cite{King2021,Golub2021,Lei2021,Kolodzeiski2023,Sayfutyarova2017,Zou2020,Bao2019}.

The most straightforward approach is to select specific orbitals from the preceding Hartree-Fock calculations.
The \verb$.sort_mo()$ method of the MCSCF classes in PySCF (including CASCI and CASSCF and the derived state-average MCSCF classes)
provides the basic functionalities for this task.
Another common strategy for the MCSCF initial guess is the use of natural orbitals from UKS or MP2 calculations.
For unrestricted Kohn-Sham (UKS) calculations, the spin-traced density matrix can be computed.
The SciPy diagonalization routine can then be used to obtain the natural orbitals:
\begin{verbatim}
mf = mol.UKS().run()
dma, dmb = mf.make_rdm1()
dm_sf = dma + dmb
nat_occ, nat_orb = scipy.linalg.eigh(
        dm_sf, mf.get_ovlp(), type=2)
nat_occ = nat_occ[::-1]
nat_orb = nat_orb[:,::-1]
\end{verbatim}
For MP2 calculations, it should be noted that the MP2 density matrix is computed by default in the molecular orbitals (MO) representation.
When calling the \verb$make_rdm1()$ method of the MP2 classes, the option
\verb$ao_repr=True$ should be specified to ensure that the density matrix is
output in the AO representation.
The subsequent steps for obtaining the natural orbitals are the same as the UKS case.

More sophisticated orbital initial guess can be constructed using the PySCF built-in modules \verb$APC$, \verb$AVAS$, and \verb$dmet_cas$.
The APC (approximate pair coefficient) method\cite{King2021} selects active space based on
the order of APC entropy, which is a measure of the interaction between the virtual
orbitals and the occupied orbitals based on the exchange matrix.
The following code example can be utilized to generate active space as well as the initial orbitals:
\begin{verbatim}
myapc = apc.APC(mf, max_size=12)
ncas, nelecas, mo_init_guess = myapc.kernel()
mc = mf.CASSCF(ncas, nelecas)
mc.run(mo_coeff=mo_init_guess)
\end{verbatim}

This APC initial guess remains the use of canonical mean-field orbitals.
When handling transition metal complexes, the active center are typically chosen to the $d$-shell of the transition metal.
One would like to construct the active space to include the atomic characteristic orbitals of the active transition metal.
In such cases, the active space could be constructed using the orbitals from the transition metal, supplemented by ligand orbitals
based on their entanglement with the orbitals of the transition metal.
The density-matrix embedding theory (DMET) based scheme, including the AVAS
(atomic valence active space)\cite{Sayfutyarova2017} and \verb$dmet_cas$ were developed to address this need.
The AVAS and \verb$dmet_cas$ selection schemes are largely interchangeable, with minor difference in the treatment of open-shell electrons.
Here, we demonstrate the use of \verb$dmet_cas$ for this category:
\begin{widetext}
\begin{lstlisting}[style=mystyle]
ncas, nelecas, mo_init_guess = dmet_cas.guess_cas(mf, mf.make_rdm1(), 'Fe 3d')
mc = mf.CASSCF(ncas, nelecas)
mc.run(mo_coeff=mo_init_guess)
\end{lstlisting}
\end{widetext}
The \verb$dmet_cas$ function requires to specify the orbitals of interest using their atomic labels, such as \verb$"Fe 3d"$,
as the primary components of the active space.
This function uses these labels to identify atomic orbitals (using the \verb$mol.search_ao_label()$ method).
It then decomposes the density matrix with respect to these atomic orbitals to extract their entangled bath orbitals.
The specified atomic orbitals along with the entangled bath form the active space.

\subsubsection{Initial Guess for Excited States}
By default, the initial guess for excited states is constructed based on the energy gap between occupied and virtual orbitals.
To solve $N$ excited states, the $N$ single orbital-pair excitations with the smallest occupied-virtual energy differences are selected as the initial guess.

A common issue with this initial guess construction is that it may overlook certain low-lying excited states.
The initial $N$ states might not include the symmetry sectors of these low-lying excited states.
Consequently, these states are completely excluded from the subspace spanned by
the Davidson diagonalization iterations, and thus, they are not represented in the results.

One effective approach to address this problem is to compute more states than the required number,
hoping that the enlarged initial space could potentially include more symmetry sectors.
However, this approach increases the computational costs, particularly for the CI and equation-of-motion CC (EOM-CC) computations.
These methods demand substantial CPU resources, memory usage, and I/O storage, all of which scale proportionally with the number of states being computed.

To reduce the computation costs, we can convert the TDDFT results to the
initial guess for these methods, assuming the TDDFT is more cost effective.
The example below demonstrates how to first perform a singlet TDDFT computation and then use the TDDFT results as the initial guess for a singlet type of EOM-CC-CCSD.
The initial guess for EOM-CC consists of a flattened vector containing single and double excitation coefficients.
Only the $X$ amplitudes from the TDDFT results are used to replace the single excitation coefficients.
\begin{widetext}
\begin{lstlisting}[style=mystyle]
td = mf.TDA().run(nstates=10)
cc = mf.CCSD().run()
eom = cc.EOMEESinglet()
n = eom.vector_size()
nocc, nmo = cc.nocc, cc.nmo
nvir = nmo - nocc
r2 = np.zeros((nocc, nocc, nvir, nvir))
nstates = 3
guess = [eom.amplitudes_to_vector(x, r2) for x, y in td.xy[:nstates]]
eom.kernel(nstates, guess=guess)
\end{lstlisting}
\end{widetext}

Both TDDFT and EOM-CC-CCSD in PySCF support singlet, triplet, and spin-flip computations.
Similar transformations can be applied to triplet and spin-flip computations as well.
Details for these conversions are not demonstrated here.

\subsection{Second-Order Self-Consistent Field (SOSCF) Convergence}
\label{sec:soscf}
The second order SCF (SOSCF) algorithm in PySCF\cite{Sun2017} is an exact second-order solver that computes the orbital Hessian precisely.
Originally, it was not designed to reduce the computation timing for arbitrary inputs.
Instead, it focuses on ensuring convergence to a local minimum solution that is close to the provided initial guess.
Unlike the quasi-newton implementations in other packages\cite{Neese2000,Thoegersen2005,Fischer1992,Slattery2024},
it does not employ aggressive optimization techniques, such as line search or
extensive trust region strategies, to reduce the number of iterations.

For simple systems with large orbital energy gaps, SOSCF may not necessarily outperform the default DIIS (direct inversion in the iterative subspace) scheme.
However, as an exact second-order solver, it exhibits advantages by achieving
quadratic convergence when the initial guesses that are close to a local minimum.
For challenging systems, such as transition metal complexes, open-shell systems, dissociated molecules, or systems with small HOMO-LUMO gaps,
switching to the SOSCF solver after several DIIS iterations often yields better performance than using the DIIS algorithm throughout the entire convergence process.

To enable an SOSCF calculation, we can invoke the \verb$newton$ method to create an SOSCF instance.
Please note that the SOSCF instance is independent from the original mean-field instance.
Solving the mean-field wavefunction using the SOSCF instance does not alter the original mean-field object (due to the side-effect free design of PySCF).
\begin{verbatim}
# This mf object utilizes DIIS algorithm
mf = mol.UKS(xc='pbe')
# Create an independent SOSCF object
soscf_mf = mf.newton()
soscf_mf = soscf_mf.run()
mf = soscf_mf.undo_newton()
\end{verbatim}
The \verb$undo_newton()$ method, called at the end,
reverts the SOSCF instance back to the regular mean-field instance.
This conversion can be omitted if the mean-field instance is subsequently used for post-HF calculations.
This conversion is automatically applied within the post-HF methods in PySCF.

Given that the SOSCF method in PySCF is not optimized for calculations starting
from poor initial orbitals, it is advisable to perform a fast preliminary computation first.
This preliminary step could be performed with approximate integrals (Section~\ref{sec:approx:integrals}), different functionals, or even a different basis set.
When a different basis set is used, the density matrix from the preliminary
computation should be projected, as discussed in
Section~\ref{sec:mf:init:guess}, and then supplied to the SOSCF kernel function.

\begin{figure}
  \begin{center}
    \includegraphics[width=0.5\textwidth]{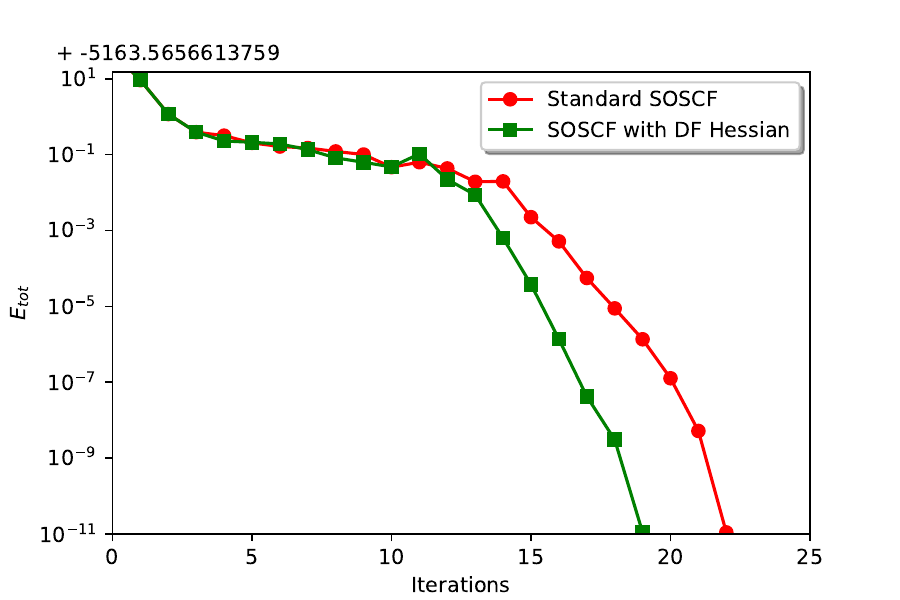}
  \caption{Convergence rate comparison between standard SOSCF and SOSCF with approximate DF orbital Hessian.}
  \label{fig:df:soscf}
  \end{center}
\end{figure}

The SOSCF method depends on the orbital Hessian, which is the most computationally demanding operation in the SOSCF process.
PySCF allows the use of approximate integrals for the orbital
Hessian, while the total energy and orbital gradients are evaluated using exact integrals.
Although the approximate orbital Hessian may slightly affect the
convergence performance, the final converged results remain unchanged, as the total
energy and orbital gradients are computed accurately.
The following example demonstrates the use of density fitting integrals to
approximate the orbital Hessian in SOSCF.
\begin{verbatim}
mf = mol.UKS(xc='pbe0')
mf = mf.newton()
# DF approximate integrals for orbital Hessian
mf = mf.density_fit()
mf = mf.run()
\end{verbatim}
Figure~\ref{fig:df:soscf} illustrates the convergence rate of the SOSCF for a challenging system (the
iron-sulfur molecule in Figure~\ref{fig:iron:sulfur}).
The calculations were performed using the PBE0 functional with the def2-TZVP basis set.
The initial guess for this test was constructed from the spin-polarized density matrix, as shown in Figure~\ref{fig:iron:sulfur}.
This was not an optimal initial guess for this particular case.
Initially, the decrease in energy is slow.
As the solver approaches the local minimum, it exhibits rapid convergence.
The auxiliary basis set (ABS) def2-universal-jfit\cite{Weigend2008} is used in the DF approximate orbital Hessian.
This ABS is not considered suitable for standard DFT calculations when using hybrid functionals.
Despite this, the DF approximate orbital Hessians performs exceptionally well.
The convergence speed of SOSCF using the approximate orbital Hessian and the
exact Hessians is essentially identical.\cite{Sun2017}

\subsection{Approximate Integral Computation}
\label{sec:approx:integrals}
In many quantum chemistry calculations, particularly the mean-field level theory, integral computation dominates the computation works.
Altering the integral computation algorithm could improve computation efficiency.
PySCF offers several integral algorithms to accelerate the evaluation of exchange-correlation (XC) functional numerical integration, the Coulomb matrix, and the HF exchange matrix.

\subsubsection{Multigrid Integration}
The numerical integration of XC functionals can be significantly accelerated using the multigrid integration algorithm.
The multigrid algorithm for XC functional integration was originally developed by the CP2K program package\cite{Kuehne2020}.
This algorithm was introduced into PySCF 1.6 with adaptations to the PySCF code structure and enhanced control of numerical precision.
Recently, the algorithm has been carefully optimized and migrated to the GPU architecture, and is currently under development in the GPU4PySCF package\cite{Li2025b}.
The advantages of the multigrid algorithm, including its recent developments,
will be further detailed in upcoming publications\cite{Zhang2025}.

The multigrid algorithm leverages the locality of Gaussian basis functions to reduce computational efforts in numerical integration.
It is specifically tailored for uniform integration grids.
Uniform grids are not as effective as Becke atomic grids in handling core electrons near the nuclei.
Very dense grids are required to accurately describe the electron density of core electrons.
However, these dense grids can lead to exceptionally large memory usage, and a very high cost for the discrete Fourier transforms that the multigrid algorithm relies on.
Except for light elements, the multigrid algorithm is generally not used directly for all-electron calculations.
Rather, it is typically combined with pseudo-potentials to screen the core electrons.

The PySCF periodic boundary condition (PBC) DFT program employs uniform grids for integration evaluations.
The multigrid can be naturally applied to PBC computations.
However, the multigrid program is not compatible with the default Becke grids scheme used by the PySCF molecular DFT code.
Additional setup steps are necessary for molecular systems.
The example below illustrates how to configure the multigrid algorithm for a molecule cluster:
\begin{widetext}
\begin{lstlisting}[style=mystyle]
# 1. Mole object for the water cluster. Pseudo potentials are applied to screen core electrons.
mol = pyscf.M(atom='xyz-of-cluster', basis='gth-tzvp', pseudo='gth-pbe')
# 2. Define a box large enough to include all electron density of the molecule
box = np.eye(3) * 30
# 3. Convert the Mole object to a Cell object. Setting dimension=0 to screen
# Coulomb interactions between periodic images
cell = mol.to_cell(box=box, dimension=0)
# 4. Replace the default numerical integrator with the MultiGridNumInt class for DFT instance
mf = cell.RKS(xc='pbe')
mf._numint = multigrid.MultiGridNumInt(cell)
# 5. Adjust mesh to refine the size of uniform grids.
mf._numint.mesh = [105, 105, 105]
mf.run()
\end{lstlisting}
\end{widetext}

In this code snippet, we intentionally demonstrates how to convert between a Mole instance and a Cell instance using the \verb$.to_cell()$ method.
However, the first three steps can be combined by directly instantiating a \verb$Cell$ class with the box and the \verb$dimension=0$ setting.
To ensure that the PBC-based DFT computation numerically matches the molecular DFT results, the \verb$dimension=0$ setting is necessary.

Adjusting the multigrid mesh can reduce computation costs, though it results in a trade-off in integration accuracy.
During each step of the SCF iterations, the package can display the number of electrons calculated from the multigrid integration.
This output can be used as an indicator of the error introduced by the truncated mesh.

Although the multigrid may introduce certain errors in the description of core
electrons, it remains a valuable tool for accelerating all-electron DFT simulations in various ways.
The multigrid can be used for preliminary computations to generate an initial guess.
When combined with the Second-Order Self-Consistent Field (SOSCF) method, it can be used to approximate the orbital Hessian.
The code snippet below shows the use of the multigrid algorithm with the SOSCF method.
\begin{verbatim}
mf = mol.RKS(xc='pbe')
cell = mf.to_cell(box, dimension=0)
mf = mf.newton()
mf._numint = multigrid.MultiGridNumInt(cell)
mf._numint.mesh = [105, 105, 105]
mf.run()
\end{verbatim}

\subsubsection{Density fitting}
The density fitting (DF) method, also known as the resolution-of-identity (RI)
method, employs a three-index tensor to approximate four-center two-electron
integrals\cite{Vahtras1993}
\begin{widetext}
\begin{equation}
  (\mu\nu|\kappa\lambda)
  = \int \chi_\mu(\mathbf{r}_1) \chi_\nu(\mathbf{r}_1) \frac{1}{r_{12}}
  \chi_\kappa(\mathbf{r}_2) \chi_\lambda(\mathbf{r}_2) d \mathbf{r}_1 d \mathbf{r}_2
  \approx \sum_P T_{P,\mu\nu} T_{P,\kappa\lambda}
  \label{eq:df:eri}
\end{equation}
\end{widetext}
Here, $\chi(\mathbf{r})$ denotes the orbital basis functions.
The index $P$ corresponds to the auxiliary basis functions $\chi_P(\mathbf{r})$ which are utilized to construct the $T$ tensor,
\begin{align}
  M_{PQ} &= (\chi_P|\chi_Q) = L L^T,
  \\
  T_{P,\mu\nu} &= \sum_{Q} (L^{-1})_{PQ} (\chi_Q|\mu\nu) .
\end{align}

For small or medium-sized molecules, using this integral approximation has several advantages:
\begin{itemize}
\item It is more efficient than the traditional analytical method of computing four-index integrals.
\item It requires less memory storage and it reduces I/O pressure.
\item It offers reduced scaling in the AO to MO integral transformation.
\end{itemize}
As a result, the DF approximation significantly enhances the performance in the
computation of Hartree-Fock exchange and various electron-correlation methods.
It has been adopted as the default in many software packages (such as Psi4, ORCA, Bagel).
However, PySCF does not set it as the default computation scheme.

In PySCF, the DF approximation can be manually enabled for various methods, including Hartree-Fock, DFT, MCSCF, MP2, and CCSD.
For mean-field methods, the \verb$.density_fit()$ method can be used to create DF mean-field instances.
Based on a DF mean-field instance, calling its shortcut methods such as
\verb$.CASCI()$ and \verb$.CASSCF()$ will inherit the DF approximation in these subsequent post-HF calculations.
However, as of PySCF 2.9, the package does not provide similar shortcuts for other post-HF methods.
Enabling DF for MP2 and CC methods requires explicitly importing the necessary Python modules:
\begin{verbatim}
from pyscf.mp import dfmp2, dfump2
from pyscf.cc import dfccsd, dfuccsd
mol.RHF().density_fit().run().DFMP2()
mol.UHF().density_fit().run().DFMP2()
mol.RHF().density_fit().run().DFCCSD().run()
mol.UHF().density_fit().run().DFCCSD().run()
\end{verbatim}

Besides the post-HF methods, DF can also be used to approximate the orbital Hessian for SOSCF or MCSCF calculations.
Although the orbital Hessian is approximated, the energies are still evaluated using the standard integrals.
To apply the orbital Hessian for SOSCF, one simply needs to call the \verb$.density_fit()$ method of an SOSCF instance, as shown previously (Section~\ref{sec:soscf}).
For the CASSCF or state-average CASSCF methods, their \verb$.approx_hessian()$ method can be called to approximate the orbital Hessian.

The computational cost and accuracy of the DF approximation are significantly
influenced by the quality of the auxiliary basis sets (ABS)\cite{Hill2013,Weigend2008,Eichkorn1995,Eichkorn1997}.
Despite the widespread application of DF approximations, the availability of pre-optimized ABS specifically for DF remains limited.
For many orbital basis sets (OBS) and quantum chemistry methods, the lack of specifically optimized ABS is a common issue.
To address this issue, several automatic ABS generation schemes have been
developed\cite{Aquilante2009,Lehtola2021,Lehtola2023,Stoychev2017,Aquilante2007,Yang2007,Diaz-Tinoco2025}.
Schemes like AutoAux\cite{Stoychev2017} and AutoABS\cite{Yang2007} are available in the PySCF package.
Additionally, PySCF maintains a cache of ABS and OBS combinations recommended by
the Psi4 package\cite{psi42025basissets}.
When the OBS for a particular element is not found in the recommendation
cache, PySCF employs an internal ABS generation program to generate
even-tempered Gaussian (ETG) basis sets.

In the three-center integrals $(\mu\nu|\chi_P)$, we denote the angular momentum of the two OBS as $l_\mu$ and $l_\nu$, and the angular momentum of the ETG ABS as $L$.
The exponent of the Gaussian function within the OBS is represented as
$\alpha_l$, and the exponent of the auxiliary ETG as $\alpha_L$.
For each angular momentum $L = l_\mu + l_\nu$, the PySCF internal ABS generation program determines the minimum exponent
\begin{equation}
  \alpha_{L,\mathrm{min}} = \sqrt{\alpha_{l_\mu,\mathrm{min}} \cdot \alpha_{l_\nu,\mathrm{min}}},
\end{equation}
and the maximum exponent
\begin{equation}
  \alpha_{L,\mathrm{max}} = \sqrt{\alpha_{l_\mu,\mathrm{max}} \cdot \alpha_{l_\nu,\mathrm{max}}}.
\end{equation}
Here, $\alpha_{l}^\mathrm{max}$ and $\alpha_{l}^\mathrm{min}$ are the maximum and
minimum exponents for each angular momentum $l$ of the Gaussian function within the OBS.
The exponents for the auxiliary ETG are then generated as
\begin{equation}
\alpha_{L,\mathrm{min}} \cdot 2^n, \quad n = 0, \ldots, m
\end{equation}
The upper bound $m$ is chosen such that $\alpha_{L,\mathrm{min}} \cdot 2^m < \alpha_{L,\mathrm{max}}$.

\begin{table}[htp]
\centering
\caption{DF approximation errors for various type of electron repulsion
integrals (in Hartree) for the Fe atom in the def2-QZVP basis set.}
\label{tab:df:eri:errors}
\begin{tabular}{llllllll}
\hline
           &  universal-jkfit     & AutoABS             & AutoAux             & PySCF built-in     \\
\hline
$N_{aux}$  &  264                 & 270                 & 550                 & 484                \\
$(ss|ss)$  &  3.77$\times10^{-4}$ & 0.32$\times10^{-4}$ & 0.25$\times10^{-5}$ & 2.06$\times10^{-5}$\\
$(sp|sp)$  &  2.90$\times10^{-4}$ & 2.36$\times10^{-4}$ & 0.32$\times10^{-5}$ & 2.69$\times10^{-5}$\\
$(sd|sd)$  &  2.17$\times10^{-4}$ & 7.59$\times10^{-4}$ & 0.98$\times10^{-5}$ & 9.27$\times10^{-5}$\\
$(sf|sf)$  &  1.58$\times10^{-3}$ & 11.0$\times10^{-3}$ & 0.45$\times10^{-4}$ & 3.61$\times10^{-4}$\\
$(sg|sg)$  &  1.48$\times10^{-3}$ & 0.90$\times10^{-3}$ & 0.31$\times10^{-4}$ & 2.86$\times10^{-4}$\\
$(pp|pp)$  &  8.54$\times10^{-3}$ & 2.68$\times10^{-3}$ & 1.06$\times10^{-4}$ & 9.91$\times10^{-4}$\\
$(pd|pd)$  &  0.12$\times10^{-2}$ & 6.61$\times10^{-2}$ & 1.20$\times10^{-4}$ & 6.80$\times10^{-4}$\\
$(pf|pf)$  &  2.04$\times10^{-2}$ & 7.40$\times10^{-2}$ & 1.14$\times10^{-4}$ & 8.10$\times10^{-4}$\\
$(pg|pg)$  &  1.18$\times10^{-2}$ & 0.516               & 1.80$\times10^{-3}$ & 0.67$\times10^{-3}$\\
$(dd|dd)$  &  0.140               & 0.0809              & 0.28$\times10^{-3}$ & 2.03$\times10^{-3}$\\
$(df|df)$  &  0.063               & 2.046               & 1.05$\times10^{-2}$ & 0.34$\times10^{-2}$\\
$(dg|dg)$  &  0.195               & 0.671               & 1.39$\times10^{-2}$ & 0.35$\times10^{-2}$\\
$(ff|ff)$  &  1.005               & 2.018               & 3.88$\times10^{-2}$ & 3.77$\times10^{-2}$\\
$(fg|fg)$  &  0.685               & 1.267               & 0.657               & 0.6567             \\
$(gg|gg)$  &  0.566               & 0.838               & 0.458               & 0.4564             \\
\hline
total      &  3.684               & 12.24               & 1.866               & 1.830              \\
\hline
\end{tabular}
\end{table}

Please note that the built-in ABS generation scheme was initially programmed to
to provide a set of ABS when optimized ABS are not available.
However, this scheme is not specifically optimized. 
The generated ETG basis sets lack functions with large exponents, which may lead to insufficient Coulomb repulsion between core electrons.
Consequently, the electrons experience stronger nuclear attraction, which typically results in a slightly lower electronic energy.
In Table \ref{tab:df:eri:errors}, we display a comparison of errors for the DF
approximation using the def2-universal-jkfit\cite{Weigend2008}
basis sets, AutoABS, AutoAux, and the PySCF built-in ETG basis sets.
The error data shows the differences between the diagonals of the exact integrals and
the DF approximate integrals, calculated using the method described by Lehtola\cite{Lehtola2023}
\begin{equation}
  \Delta = \sum_{\mu\nu}\left[(\mu\nu|\mu\nu) - \sum_P T_{P,\mu\nu} T_{P,\mu\nu}\right]
\end{equation}
The built-in ETG of PySCF is more precise than smaller basis sets such as def2-universal-jkfit or AutoABS.
When compared to AutoAux basis sets, they are generally comparable.
The PySCF ETG tends to exhibit larger errors.
For integrals involving high angular momentum OBS, the accuracy of PySCF ETG is close to or even better than that of AutoAux.

The original ABS generation program is retained in PySCF to maintain backward compatibility.
To override the PySCF default ABS management in PySCF, we can assign
specifically generated ABS through the keyword argument \verb$auxbasis$ when initializing DF methods.
For example, the ABS produced by the \verb$autoaux$ module can be assigned to DF-DFT calculations as follows:
\begin{verbatim}
from pyscf.df import autoaux
mf = mol.RKS(xc='pbe0').density_fit(
        auxbasis=autoaux(mol))
\end{verbatim}

For larger molecular systems, the DF approximation can become a bottleneck due to
computational costs and the storage of the intermediate three-index tensor.
At the mean-field level theory, a long-range density fitting (LRDF)
scheme\cite{Sun2023} was developed in PySCF to address these challenges.
The full-range Coulomb potential can always be written as the summation of the long-range
(LR) potential and the short-range (SR) potential
\begin{equation}
  \frac{1}{r_{12}}
  = \underbrace{\frac{\mathrm{erf}(\omega r)}{r_{12}}}_{\text{LR part}}
  + \underbrace{\frac{\mathrm{erfc}(\omega r)}{r_{12}}}_{\text{SR part}}.
\end{equation}
The LRDF scheme employs the density fitting technique to approximate the LR Coulomb
integrals, while the SR Coulomb integrals are computed exactly:
\begin{equation}
  (\mu\nu|\kappa\lambda)
  \approx (\mu\nu|\kappa\lambda)^\mathrm{SR}
  + \sum_P T_{P,\mu\nu}^\mathrm{LR} T^\mathrm{LR}_{P,\kappa\lambda}.
  \label{eq:lrdf}
\end{equation}
The LR three-index tensor $T^\mathrm{LR}$ is evaluated with the LR Coulomb metric
\begin{align}
  M_{PQ} &= (\chi_P|\chi_Q)^\mathrm{LR} = L L^T,
  \\
  T_{P,\mu\nu} &= \sum_{Q} (L^{-1})_{PQ} (\chi_Q|\mu\nu)^\mathrm{LR} .
\end{align}
The computation of the short-range (SR) integrals in Eq. \eqref{eq:lrdf} scales linearly with system size.
Compared to the full-range Coulomb potential, the rank of the LR integral tensor is substantially lower.
One can use a small auxiliary basis set for the LRDF tensor to achieve sufficient accuracy.
The size of the auxiliary dimension needed is approximately only 1/10 of the size of the orbital basis functions.
This leads to a significant reduction in computational resources and storage.
The code example below demonstrates how to apply the LRDF approximation.
\begin{verbatim}
from pyscf.lrdf import LRDF
mf = mol.RKS(xc='pbe0')
mf.with_df = LRDF(mol)
\end{verbatim}
In PySCF 2.9, the LRDF module is available in the PySCF-Forge repository.
The PySCF-Forge package must be installed (\verb$pip install pyscf-forge$) before running this example.

\section{Leveraging Modern Python Tools}
\label{sec:python:tools}
The Python community has developed numerous useful tools for scientific programming and numerical computation.
Integrating PySCF with these tools can significantly improve the efficiency of
theoretical method development and computational simulations.
This section presents the use of just-in-time (JIT) compilation, automatic
differentiation (AD) to enhance the productivity of PySCF.

\subsection{Just-In-Time (JIT) Compilation}
\label{sec:python:jit}
Numba JIT compilation can significantly accelerate Python code by translating Python functions to optimized machine code at runtime.
It can be integrated with PySCF functionality to quickly implement features that are not yet available in the package.
To illustrate this capability, we will examine a feature request from a PySCF GitHub issue as an example.

\begin{widetext}
This request calls for a program to compute the overlap integrals between the absolute values of two orbitals.
\begin{equation}
  |S|_{\mu\nu} = \int |\chi_\mu(\mathbf{r})| |\chi_\nu(\mathbf{r})| d \mathbf{r}
\end{equation}
For Cartesian Gaussian-type orbitals (GTO),
this integral can be decomposed into the product of three Cartesian components:
\begin{equation}
  |S|_{\mu\nu} = I_x \cdot I_y \cdot I_z.
\end{equation}
Each component is a Gaussian type overlap integral
\begin{equation}
  I_x = \int |(x-X_\mu)^{m_x} \exp(-\alpha_\mu (x-X_\mu)^2)|\cdot|(x-X_\nu)^{n_x} \exp(-\alpha_\nu (x-X_\nu)^2)| dx.
\end{equation}
Given that the exponential component is always positive, the integrands can be
rewritten as
\begin{equation}
  |(x-X_\mu)^{m_x} (x-X_\nu)^{n_x}| \exp(-\alpha_\mu (x-X_\mu)^2-\alpha_\nu (x-X_\nu)^2)
\end{equation}
Employing the Gaussian product theorem, we can simplify the product of two Gaussian functions
\begin{equation}
  \exp\left(-\alpha_\mu (x-X_\mu)^2-\alpha_\nu (x-X_\nu)^2\right)
  = \exp\left(-\theta_{\mu\nu}
  (X_\mu-X_\nu)^2\right)
  \exp(-\left(\alpha_\mu+\alpha_\nu)(x-X_p)^2\right),
\end{equation}
where
\begin{align}
  \theta_{\mu\nu} &= \frac{\alpha_\mu\alpha_\nu}{\alpha_\mu+\alpha_\nu},
  \\
  X_p &= \frac{\alpha_\mu X_\mu + \alpha_\nu X_\nu}{\alpha_\mu + \alpha_\nu}.
\end{align}
\end{widetext}
Although $I_x$ appears very similar to the standard Gaussian integral, unless the
exponents $m_x$ and $n_x$ are both even, it cannot be analytically computed due to the absolute value of the polynomial.
Noting that the integral retains the form of a Gaussian integral, we can approximate it using Gauss-Hermite quadrature, leading to
\begin{equation}
  I_x \approx \exp(-\theta_{\mu\nu}(X_\mu-X_\nu)^2)\sum_n w_n |(x_n-X_\mu)^{m_x} (x_n-X_\nu)^{n_x}| .
\end{equation}
Here, $x_n$ and $w_n$ are transformed from the standard roots $u$ and weights
$\omega$ of Gauss-Hermite quadrature
\begin{align}
  x_n &= \frac{u_n}{\sqrt{\alpha_\mu+\alpha_\nu}} + X_p,
  \\
  w_n &= \frac{\omega_n}{\sqrt{\alpha_\mu+\alpha_\nu}}.
\end{align}
When either $m_x$ or $n_x$ is odd, the polynomial part is not an analytical
function and cannot be represented using a finite number of polynomials.
To minimize the integral error, a relatively large number of quadratures is required.

The integral code can be easily implemented in Python, as demonstrated in the following example:
\begin{widetext}
\begin{lstlisting}[style=mystyle]
@numba.njit(fastmath=True, cache=True)
def primitive_overlap(li, lj, ai, aj, ci, cj, Ra, Rb, roots, weights):
    norm_fac = ci * cj
    # Unconventional normalization for Cartesian functions in PySCF
    if li <= 1: norm_fac *= ((li*2+1)/(4*np.pi))**.5
    if lj <= 1: norm_fac *= ((lj*2+1)/(4*np.pi))**.5
    aij = ai + aj
    Rab = Ra - Rb
    Rp = (ai * Ra + aj * Rb) / aij
    theta_ij = ai * aj / aij
    scale = 1./np.sqrt(aij)
    norm_fac *= scale**3 * np.exp(-theta_ij * Rab.dot(Rab))
    x = roots * scale + Rp[:,None]
    xa = x - Ra[:,None]
    xb = x - Rb[:,None]
    mu = xa ** np.arange(li+1)[:,None,None]
    nu = xb ** np.arange(lj+1)[:,None,None]

    nfi = (li+1)*(li+2)//2
    nfj = (lj+1)*(lj+2)//2
    s = np.empty((nfi, nfj))
    i = 0
    for ix in range(li, -1, -1):
        for iy in range(li-ix, -1, -1):
            iz = li - ix - iy
            j = 0
            for jx in range(lj, -1, -1):
                for jy in range(lj-jx, -1, -1):
                    jz = lj - jx - jy
                    Ix = (mu[ix,0] * nu[jx,0] * weights).sum()
                    Iy = (mu[iy,1] * nu[jy,1] * weights).sum()
                    Iz = (mu[iz,2] * nu[jz,2] * weights).sum()
                    s[i,j] = Ix * Iy * Iz * norm_fac
                    j += 1
            i += 1
    return s
\end{lstlisting}
\end{widetext}

The numerical computation workload is intensive in this function.
Although NumPy functions already offer a significant speedup compared to plain
Python, they are not able to fully utilize CPU capabilities, particularly in terms of using the instruction-level parallelism instructions provided by CPU.
Using Numba JIT compilation can accelerate the numerical computation in this integral code.

To generate efficient machine code with Numba JIT, we have enabled the {\it no-python} mode
(\verb$@numba.njit$) in this code example.
While the {\it no-python} mode can produce more efficient machine code for
numerical computations,
it restricts the use of Python data structures and complex Pythonic features.
We are unable to use Python classes or Pythonic iterators within this decorator.
Loops and computation expressions have to be written in a straightforward and plain style.
In other words, the basis information provided by the PySCF \verb$Mole$ class cannot be directly accessed.
They have to be converted into simple data representation before being passed the Numba JIT compiled function.
This is an inconvenient consequence of using Numba JIT compilation.

The {\it no-python} compilation mode can leverage more specialized SIMD instructions which might not be supported by NumPy precompiled libraries.
This mode automatically employs broader SIMD vectorization, allowing the CPU to execute multiple floating-point operations within a single CPU cycle.
This capability is facilitated by the underlying LLVM compiler, which makes Numba JIT-compiled numerical computations faster than NumPy operations in various scenarios.
In the \verb$primitive_overlap$ function, we have additionally enabled the
\verb$fastmath=True$ option in the \verb$njit$ decorator to activate the FMA (fused multiplication-add) SIMD instructions.
The FMA instruction can perfrom an addition and a multiplication in a single CPU cycle, potentially doubling the performance.
However, using FMA can lead to discrepancies compared to performing the multiplication and addition separately\cite{Boldo2011,Muller2010}.
Therefore, it is disabled by default in Numba JIT.

Numba JIT automatically expands NumPy Ufuncs (universal functions) or broadcast operations into loops.
However, manually rewriting some of these operations into explicit loops can still benefit computational efficiency.
For instance, the statement
\begin{verbatim}
mu = xa ** np.arange(li+1)[:,None,None]
\end{verbatim}
can be manually unrolled as
\begin{lstlisting}[style=mystyle]
for n in range(nroots):
    powx = 1.
    powy = 1.
    powz = 1.
    mu[0,0,n] = 1.
    mu[0,1,n] = 1.
    mu[0,2,n] = 1.
    for i in range(1, li+1):
        powx = powx * xa[0,n]
        powy = powy * xa[1,n]
        powz = powz * xa[2,n]
        mu[i,0,n] = powx
        mu[i,1,n] = powy
        mu[i,2,n] = powz
\end{lstlisting}
This unrolled version reduces the number of expensive power operations, the number of memory accesses, and the overhead introduced by loops.
When calculating \verb$Ix$, \verb$Iy$, \verb$Iz$,
the three Ufunc statements are expanded into three loop blocks.
We can further optimize this part of the Numba code to reduce the loop overhead and memory footprint, enhancing the overall performance of the simulation.
\begin{lstlisting}[style=mystyle]
Ix = 0
Iy = 0
Iz = 0
for n in range(nroots):
    Ix += mu[ix,0,n]*nu[jx,0,n]*weights[n]
    Iy += mu[iy,1,n]*nu[jy,1,n]*weights[n]
    Iz += mu[iz,2,n]*nu[jz,2,n]*weights[n]
\end{lstlisting}

There are several techniques to enhance the performance of Numba JIT.
We will not describe each of these optimization techniques in detail in this paper.
For a comprehensive discussion, please refer to the dedicated chapters in the
book {\it Python for Quantum Chemistry}\cite{Sun2025a}.

Subsequently, we invoke the PySCF APIs to gather the necessary information of the
GTO functions for this integral function, including angular momentum,
exponents, normalization coefficients, and locations.
We then iterate over all pairs of GTOs,
supplying these data to the integral evaluation function and assemble the output into a matrix:
\begin{widetext}
\begin{lstlisting}[style=mystyle]
@numba.njit(cache=True)
def primitive_overlap_matrix(
        ls, exps, norm_coef, bas_coords, roots, weights):
    nbas = len(ls)
    dims = [(l + 1) * (l + 2) // 2 for l in ls]
    nao = sum(dims)
    smat = np.empty((nao, nao))
    i0 = 0
    for i in range(nbas):
        j0 = 0
        for j in range(i+1):
            s = primitive_overlap(
                  ls[i], ls[j], exps[i], exps[j],
                  norm_coef[i], norm_coef[j],
                  bas_coords[i], bas_coords[j], roots, weights)
            smat[i0:i0+dims[i], j0:j0+dims[j]] = s
            # smat is a symmetric matrix
            if i != j: smat[j0:j0+dims[j], i0:i0+dims[i]] = s.T
            j0 += dims[j]
        i0 += dims[i]
    return smat

def absolute_overlap_matrix(mol, nroots=500):
    from scipy.special import roots_hermite
    assert mol.cart
    # Integrals are computed using primitive GTOs. ctr_mat transforms the
    # primitive GTOs to the contracted GTOs.
    pmol, ctr_mat = mol.decontract_basis(aggregate=True)
    # Angular momentum for each shell
    ls = np.array([pmol.bas_angular(i) for i in range(pmol.nbas)])
    # need to access only one exponent for primitive gaussians
    exps = np.array([pmol.bas_exp(i)[0] for i in range(pmol.nbas)])
    # Normalization coefficients
    norm_coef = gto.gto_norm(ls, exps)
    # Position for each shell
    bas_coords = np.array([pmol.bas_coord(i) for i in range(pmol.nbas)])

    r, w = roots_hermite(nroots)
    s = primitive_overlap_matrix(ls, exps, norm_coef, bas_coords, r, w)
    return ctr_mat.T.dot(s).dot(ctr_mat)
\end{lstlisting}
\end{widetext}

\subsection{Automatic Differentiation}
\label{sec:python:ad}
In the realm of automatic differentiation (AD), JAX and
Torch are the most commonly used frameworks in Python-based applications.
These AD frameworks have been employed in several quantum chemistry-related projects\cite{Zhang2022,Kasim2022,Zhou2020,Friede2024}.
Compared to Torch, JAX offers greater flexibility, with highlights in
the automatic transitions between the forward-mode (jvp) and backward-mode (vjp) differentiation,
the implicit differentiation for fixed-point iterative implementations,
and the high-order derivatives.
These features make JAX suitable for AD in complex computational programs.

The PySCFAD package developed by Zhang\cite{Zhang2022} is a differentiable framework that integrates the AD functionalities of JAX.
It offers capabilities for differentiating basis contraction coefficients, the
exponents of Gaussian orbitals, and the coordinates of atoms for quantum
chemistry methods applicable to both molecular and periodic boundary condition (PBC) systems.
The data structure and APIs of PySCFAD are very similar to those of PySCF.
Once the standard (without differentiation) implementation is established in PySCF,
adding differentiable functionalities using PySCFAD requires minimal additional effort.
As an example, we consider the optimization of auxiliary basis functions for the
density fitting method, demonstrating how PySCFAD can be used to compute
gradients for newly developed theoretical models.

The density fitting approximation Eq. \eqref{eq:df:eri} can be rewritten as
\begin{equation}
  (\mu\nu|\kappa\lambda) \approx \sum_{PQ} (\chi_P|\mu\nu) (\mathbf{M}^{-1})_{PQ}(\chi_Q|\kappa\lambda).
\end{equation}
To enhance the accuracy of the DF approximation, we consider optimizing the
exponents $\alpha_P$ of the auxiliary GTO $\chi_P$ to minimize the error
\begin{equation}
  \sum_{\mu\nu}\left[\sum_{PQ} (\chi_P|\mu\nu) (\mathbf{M}^{-1})_{PQ}(\chi_Q|\mu\nu) - (\mu\nu|\mu\nu)\right]^2.
\end{equation}
This minimization problem is equivalent to a least-squares fitting problem.
The least-squares fitting problem can be solved using the \verb$least_squares$
function provided by the SciPy package, which requires input functions to compute the
residual
\begin{equation}
  \sigma_{\mu\nu} = (\mu\nu|\mu\nu) - C_{P,\mu\nu}(P|\mu\nu),
\end{equation}
and the Jacobian of the residual
\begin{equation}
  \frac{\partial}{\partial \alpha_P}\sigma_{\mu\nu\kappa\lambda}.
\end{equation}

Using PySCF APIs, we construct the \verb$residual$ function for a molecule instance \verb$mol$
(holding OBS) and \verb$auxmol$ (containing ABS).
\begin{widetext}
\begin{lstlisting}[style=mystyle]
def residual(mol, auxmol):
    int4c2e = mol.intor('int2e', aosym='s4').diagonal()
    int3c2e = int3c_cross(mol, auxmol, aosym='s2')
    int2c2e = auxmol.intor('int2c2e')
    return np.einsum('ip,pi->i', int3c2e, np.linalg.solve(int2c2e, int3c2e.T)) - int4c2e
\end{lstlisting}
\end{widetext}

To enable the automatic differentiation of the residual function, we make the following modifications:
\begin{enumerate}
\item Replace the \verb$mol$ and \verb$auxmol$ instances with the corresponding ones from PySCFAD;
\item Substitute \verb$numpy$ with \verb$jax.numpy$.
\end{enumerate}

The code implementation is demonstrated below.
Please note the options \verb$trace_ctr_coeff=False$ and \verb$trace_ext=False$
are used to instantiating the PySCFAD \verb$mol$ instance.
These settings disable the track of derivatives for OBS, as our focus is on
optimizing the Gaussian exponents of ABS.
For the \verb$auxmol$ instance, the \verb$trace_exp$ is enabled to track the exponent derivatives.
In PySCFAD, the exponents are accessible through the \verb$.exp$ attribute of the \verb$auxmol$ instance.
Their derivatives can be accessed via the \verb$.exp$ attribute of the output
JAX \verb$PyTree$ class\cite{Jax2025}.
We explicitly call \verb$jax.jacfwd$ in this example because the size of
orbital-pairs are typically much larger than the number of unique exponents in
ABS, yielding a tall Jacobian matrix.
This Jacobian matrix structure is well-suited for the forward-mode AD.
\begin{widetext}
\begin{lstlisting}[style=mystyle]
import jax.numpy as jnp
from pyscfad import gto
from pyscfad.df.incore import int3c_corss
mol = gto.Mole()
mol.atom = '''
O    0.    0.000    0.118
H    0.    0.755   -0.471
H    0.   -0.755   -0.471'''
mol.basis = 'cc-pvdz'
mol.build(trace_ctr_coeff=False, trace_exp=False)

auxmol = gto.Mole()
auxmol.atom = mol.atom
auxmol.basis = auxbasis
mol.build(trace_ctr_coeff=False, trace_exp=True)

def residual(auxmol):
    int4c2e = mol.intor('int2e', aosym='s4').diagonal()
    int3c2e = int3c_cross(mol, auxmol, aosym='s2')
    int2c2e = auxmol.intor('int2c2e')
    return jnp.einsum('ip,pi->i', int3c2e, jnp.linalg.solve(int2c2e, int3c2e.T)) - int4c2e

jac = jax.jacfwd(residual)(auxmol)
print(jac.exp) # A vector that stores exponents derivatives
\end{lstlisting}
\end{widetext}

The residual function and its Jacobian function establish the main framework for the least-squares minimization problem.
The SciPy \verb$least_squares$ API requires that both the residual and
the Jacobian function take a NumPy vector as the input argument and
produce a NumPy vector as the output.
We would need a few additional wrapper code to adapt to the SciPy \verb$least_squares$ API.
The complete implementation, including problem setup, initialization, adaptation
code, and the \verb$least_squares$ fitting, is demonstrated below.
\begin{widetext}
\begin{lstlisting}[style=mystyle]
import jax.numpy as jnp
from pyscf.df.incore import aux_e2
from pyscfad.gto.mole import setup_exp
from pyscfad.df.incore import int3c_cross

def setup_auxbasis(mol, auxbasis):
    int4c2e = mol.to_pyscf().intor('int2e', aosym='s4').diagonal()
    tril_idx = jnp.tril_indices(mol.nao)

    auxmol = gto.Mole()
    auxmol.atom = mol.atom
    auxmol.basis = auxbasis
    auxmol.build(trace_ctr_coeff=False)
    x0 = auxmol.exp
    _, _, env_ptr, unravel_exp = setup_exp(auxmol, return_unravel_fn=True)

    def f_residual_for_jax(auxmol):
        int3c2e = int3c_cross(mol, auxmol, aosym='s1')[tril_idx]
        int2c2e = auxmol.intor('int2c2e')
        return jnp.einsum('ip,pi->i', int3c2e, jnp.linalg.solve(int2c2e, int3c2e.T)) - int4c2e

    # The jacobian is typically a very tall matrix
    jax_jac = jax.jacfwd(f_residual_for_jax)

    def jac(x): # A wrapper to update _env and exp before calling JAX AD
        auxmol._env[env_ptr] = x
        auxmol.exp = jnp.array(x)
        return jax_jac(auxmol).exp

    def f_residual(x):
        auxmol._env[env_ptr] = x
        return f_residual_for_jax(auxmol)

    return f_residual, jac, x0, unravel_exp

if __name__ == '__main__':
    from scipy.optimize import least_squares
    auxbasis = {
        'C': [[0, [9., 1.]],
              [0, [3., 1.]],
              [0, [1., 1.]],
              [1, [3., 1.]],
              [1, [1., 1.]],
              [2, [1., 1.]]],
        'O': [[0, [9., 1.]],
              [0, [3., 1.]],
              [0, [1., 1.]],
              [1, [3., 1.]],
              [1, [1., 1.]],
              [2, [1., 1.]]]
    }
    f_residual, jac, x0, unravel_exp = setup_auxbasis(mol, auxbasis)
    result = least_squares(f_residual, x0, jac=jac, gtol=1e-6, verbose=2)
    print(unravel_exp(result.x))

\end{lstlisting}
\end{widetext}

There are several noteworthy aspects in this implementation:
\begin{itemize}
\item PySCFAD currently cannot effectively handle the permutation symmetry in the integrals.
When calculating the \verb$int3c2e$ tensor, the permutation symmetry between
orbital pairs (as indicated by \verb$aosym='s1'$) is not utilized.
Instead, we compute the entire three-index tensor and extract the unique variables located in the lower triangular part.

\item The \verb$least_squares$ optimization continuously generate new vectors of exponents.
For each new vector, before calling the residual function or the Jacobian function, we need to update the corresponding attributes of \verb$auxmol$ to incorporate the new vector.
The actual analytical Gaussian integrals are computed by the integral program in PySCF, which reads the exponents from \verb$auxmol._env$.
The assignment \verb$auxmol._env[env_ptr] = x$ synchronizes the newly generated exponents vector to \verb$auxmol._env$.
The addresses \verb$env_ptr$ here are obtained via the PySCFAD helper function \verb$setup_exp()$.
Additionally, \verb$auxmol.exp$ attribute from PySCFAD needs to be updated to post-process the integral differentiations.

\item The assignment \verb$auxmol._env[env_ptr] = x$ is an operation with side effects and is not JAX transformable.
It cannot be placed inside a JAX function.
Therefore, we only execute this assignment in the wrapper functions.
\end{itemize}

If the task is to optimize the even-tempered Gaussian parameters
$\alpha_{L,\mathrm{min}}$ and $\beta_L$,
\begin{equation}
  \alpha_{L,\mathrm{min}} \cdot \beta_L^n
\end{equation}
a similar program can implemented.
In this program, an adaptive wrapper function would need to transform the
$\alpha_{L,\mathrm{min}}$ and $\beta_L$ parameters into the PySCFAD \verb$Mole.exp$ representation.
The adaptation function can be written to comply with JAX-transformable standards, and JAX can easily handle them within the AD framework.
More detailed implementation has been documented as an example within the PySCFAD project.
For additional information on the technical aspects of JAX AD in the context of
quantum chemistry, please refer to the discussions in the book {\it Python for
Quantum Chemistry}\cite{Sun2025}.

We then use this program to examine the optimal values for the $\alpha_\mathrm{min}$ and $\beta$ parameters in the auxiliary ETG basis sets.
Table~\ref{tab:df:etg:optimize} presents the DF approximation errors for some optimized ETG basis sets for the integrals of Fe atom in the def2-TZVP OBS.
The error introduced by PySCF built-in ETG is small, implying that its size could potentially be optimized for better performance.
Reducing the size of the ETG basis set to smaller ones (ETG I and ETG II) only slightly increase the overall error.
In these ETG basis sets, the optimal $\beta$ values are between 2.0 and 2.2.
Even with a reduction of nearly half of the auxiliary basis functions (ETG III), the error remains relatively small.
However, the optimal $\beta$ value would need to be increased to 2.5 or larger.

\begin{table}[htp]
\centering
\caption{Optimized ETG parameters for DF auxiliary basis set for the Fe atom in the def2-TZVP basis set.}
\label{tab:df:etg:optimize}
\begin{tabular}{lllllllllllllll}
\hline
\multicolumn{4}{l}{PySCF built-in ETG} \\
$L$                        & 0     & 1     & 2     & 3     & 4     & 5     & 6     \\
$n_L$                     & 21    & 17    & 15    & 12    & 10    & 5     & 1     \\
$\alpha_{L,\mathrm{min}}$  & 0.084 & 0.150 & 0.124 & 0.222 & 0.182 & 0.763 & 3.196 \\
$\beta_L$                  & 2.0   & 2.0   & 2.0   & 2.0   & 2.0   & 2.0   &       \\
total error & \multicolumn{4}{l}{7.42$\times 10^{-3}$} \\
\hline
ETG I \\
$L$                       & 0     & 1     & 2     & 3     & 4     & 5     & 6     \\
$n_L$                     & 17    & 14    & 12    & 10    & 8     & 4     & 1     \\
$\alpha_{L,\mathrm{min}}$ & 0.121 & 0.188 & 0.157 & 0.237 & 0.190 & 1.152 & 3.196 \\
$\beta_L$                 & 2.200 & 1.954 & 2.202 & 1.953 & 1.937 & 2.006 &     \\
total error & \multicolumn{4}{l}{1.06$\times 10^{-2}$} \\
\hline
ETG II \\
$L$                       & 0     & 1     & 2     & 3     & 4     & 5     & 6     \\      
$n_L$                     & 14    & 11    & 10    & 8     & 6     & 4     & 1     \\
$\alpha_{L,\mathrm{min}}$ & 0.121 & 0.149 & 0.157 & 0.229 & 0.206 & 1.223 & 3.196 \\
$\beta_L$                 & 2.196 & 2.237 & 2.157 & 2.113 & 2.313 & 1.905 &       \\
total error & \multicolumn{4}{l}{2.62$\times 10^{-2}$} \\
\hline
ETG III \\
$L$                       & 0     & 1     & 2     & 3     & 4     & 5     & 6     \\
$n_L$                     & 11    & 9     & 8     & 6     & 5     & 3     & 1     \\
$\alpha_{L,\mathrm{min}}$ & 0.118 & 0.171 & 0.117 & 0.280 & 0.228 & 1.085 & 3.196 \\
$\beta_L$                 & 2.982 & 2.613 & 2.592 & 2.544 & 2.545 & 2.192 &       \\
total error & \multicolumn{4}{l}{0.139} \\
\hline
\end{tabular}
\end{table}




\section{Conclusion}
In this paper, we have provided a guide for developing and accelerating quantum chemistry program computations using the PySCF framework and modern Python programming tools.
We explored various methods to enhance computation performance by manipulating
initial guesses, applying second-order convergence, and utilizing integral
approximations, through PySCF built-in functionalities.
The PySCF extensions, GPU4PySCF and PySCFAD, further enhance the capabilities of the PySCF package.
For computationally demanding tasks, the GPU kernels provided by GPU4PySCF offer a straightforward and effective method for acceleration.
We demonstrated the use of GPU4PySCF APIs and the coordination between GPU4PySCF and PySCF features.
Additionally, Python programming tools can be integrated with the PySCF package to develop new methods and features efficiently.
One such example is the use of Numba JIT compilation.
By using this tool in conjunction with PySCF's simple APIs, we can optimize CPU resource utilization for numerical computations.
We also illustrate the process of using the PySCFAD package to develop
automatic derivatives for the exponents of auxiliary Gaussian basis functions
in density fitting methods, which are then used in basis set optimization tasks.

\section{Acknowledgments}
The author, Qiming Sun, would like to express his gratitude to Dr. Xing Zhang for helpful discussions on JAX
functionalities and for reviewing the GitHub pull request of additional AD features in PySCFAD related to this work.

\bibliography{ref}

\begin{thebibliography}{74}%
\makeatletter
\providecommand \@ifxundefined [1]{%
 \@ifx{#1\undefined}
}%
\providecommand \@ifnum [1]{%
 \ifnum #1\expandafter \@firstoftwo
 \else \expandafter \@secondoftwo
 \fi
}%
\providecommand \@ifx [1]{%
 \ifx #1\expandafter \@firstoftwo
 \else \expandafter \@secondoftwo
 \fi
}%
\providecommand \natexlab [1]{#1}%
\providecommand \enquote  [1]{``#1''}%
\providecommand \bibnamefont  [1]{#1}%
\providecommand \bibfnamefont [1]{#1}%
\providecommand \citenamefont [1]{#1}%
\providecommand \href@noop [0]{\@secondoftwo}%
\providecommand \href [0]{\begingroup \@sanitize@url \@href}%
\providecommand \@href[1]{\@@startlink{#1}\@@href}%
\providecommand \@@href[1]{\endgroup#1\@@endlink}%
\providecommand \@sanitize@url [0]{\catcode `\\12\catcode `\$12\catcode
  `\&12\catcode `\#12\catcode `\^12\catcode `\_12\catcode `\%12\relax}%
\providecommand \@@startlink[1]{}%
\providecommand \@@endlink[0]{}%
\providecommand \url  [0]{\begingroup\@sanitize@url \@url }%
\providecommand \@url [1]{\endgroup\@href {#1}{\urlprefix }}%
\providecommand \urlprefix  [0]{URL }%
\providecommand \Eprint [0]{\href }%
\providecommand \doibase [0]{http://dx.doi.org/}%
\providecommand \selectlanguage [0]{\@gobble}%
\providecommand \bibinfo  [0]{\@secondoftwo}%
\providecommand \bibfield  [0]{\@secondoftwo}%
\providecommand \translation [1]{[#1]}%
\providecommand \BibitemOpen [0]{}%
\providecommand \bibitemStop [0]{}%
\providecommand \bibitemNoStop [0]{.\EOS\space}%
\providecommand \EOS [0]{\spacefactor3000\relax}%
\providecommand \BibitemShut  [1]{\csname bibitem#1\endcsname}%
\let\auto@bib@innerbib\@empty
\bibitem [{\citenamefont {Sun}\ \emph {et~al.}(2020)\citenamefont {Sun},
  \citenamefont {Zhang}, \citenamefont {Banerjee}, \citenamefont {Bao},
  \citenamefont {Barbry}, \citenamefont {Blunt}, \citenamefont {Bogdanov},
  \citenamefont {Booth}, \citenamefont {Chen}, \citenamefont {Cui},
  \citenamefont {Eriksen}, \citenamefont {Gao}, \citenamefont {Guo},
  \citenamefont {Hermann}, \citenamefont {Hermes}, \citenamefont {Koh},
  \citenamefont {Koval}, \citenamefont {Lehtola}, \citenamefont {Li},
  \citenamefont {Liu}, \citenamefont {Mardirossian}, \citenamefont {McClain},
  \citenamefont {Motta}, \citenamefont {Mussard}, \citenamefont {Pham},
  \citenamefont {Pulkin}, \citenamefont {Purwanto}, \citenamefont {Robinson},
  \citenamefont {Ronca}, \citenamefont {Sayfutyarova}, \citenamefont
  {Scheurer}, \citenamefont {Schurkus}, \citenamefont {Smith}, \citenamefont
  {Sun}, \citenamefont {Sun}, \citenamefont {Upadhyay}, \citenamefont {Wagner},
  \citenamefont {Wang}, \citenamefont {White}, \citenamefont {Whitfield},
  \citenamefont {Williamson}, \citenamefont {Wouters}, \citenamefont {Yang},
  \citenamefont {Yu}, \citenamefont {Zhu}, \citenamefont {Berkelbach},
  \citenamefont {Sharma}, \citenamefont {Sokolov},\ and\ \citenamefont
  {Chan}}]{Sun2020}%
  \BibitemOpen
  \bibfield  {author} {\bibinfo {author} {\bibfnamefont {Q.}~\bibnamefont
  {Sun}}, \bibinfo {author} {\bibfnamefont {X.}~\bibnamefont {Zhang}}, \bibinfo
  {author} {\bibfnamefont {S.}~\bibnamefont {Banerjee}}, \bibinfo {author}
  {\bibfnamefont {P.}~\bibnamefont {Bao}}, \bibinfo {author} {\bibfnamefont
  {M.}~\bibnamefont {Barbry}}, \bibinfo {author} {\bibfnamefont {N.~S.}\
  \bibnamefont {Blunt}}, \bibinfo {author} {\bibfnamefont {N.~A.}\ \bibnamefont
  {Bogdanov}}, \bibinfo {author} {\bibfnamefont {G.~H.}\ \bibnamefont {Booth}},
  \bibinfo {author} {\bibfnamefont {J.}~\bibnamefont {Chen}}, \bibinfo {author}
  {\bibfnamefont {Z.-H.}\ \bibnamefont {Cui}}, \bibinfo {author} {\bibfnamefont
  {J.~J.}\ \bibnamefont {Eriksen}}, \bibinfo {author} {\bibfnamefont
  {Y.}~\bibnamefont {Gao}}, \bibinfo {author} {\bibfnamefont {S.}~\bibnamefont
  {Guo}}, \bibinfo {author} {\bibfnamefont {J.}~\bibnamefont {Hermann}},
  \bibinfo {author} {\bibfnamefont {M.~R.}\ \bibnamefont {Hermes}}, \bibinfo
  {author} {\bibfnamefont {K.}~\bibnamefont {Koh}}, \bibinfo {author}
  {\bibfnamefont {P.}~\bibnamefont {Koval}}, \bibinfo {author} {\bibfnamefont
  {S.}~\bibnamefont {Lehtola}}, \bibinfo {author} {\bibfnamefont
  {Z.}~\bibnamefont {Li}}, \bibinfo {author} {\bibfnamefont {J.}~\bibnamefont
  {Liu}}, \bibinfo {author} {\bibfnamefont {N.}~\bibnamefont {Mardirossian}},
  \bibinfo {author} {\bibfnamefont {J.~D.}\ \bibnamefont {McClain}}, \bibinfo
  {author} {\bibfnamefont {M.}~\bibnamefont {Motta}}, \bibinfo {author}
  {\bibfnamefont {B.}~\bibnamefont {Mussard}}, \bibinfo {author} {\bibfnamefont
  {H.~Q.}\ \bibnamefont {Pham}}, \bibinfo {author} {\bibfnamefont
  {A.}~\bibnamefont {Pulkin}}, \bibinfo {author} {\bibfnamefont
  {W.}~\bibnamefont {Purwanto}}, \bibinfo {author} {\bibfnamefont {P.~J.}\
  \bibnamefont {Robinson}}, \bibinfo {author} {\bibfnamefont {E.}~\bibnamefont
  {Ronca}}, \bibinfo {author} {\bibfnamefont {E.~R.}\ \bibnamefont
  {Sayfutyarova}}, \bibinfo {author} {\bibfnamefont {M.}~\bibnamefont
  {Scheurer}}, \bibinfo {author} {\bibfnamefont {H.~F.}\ \bibnamefont
  {Schurkus}}, \bibinfo {author} {\bibfnamefont {J.~E.~T.}\ \bibnamefont
  {Smith}}, \bibinfo {author} {\bibfnamefont {C.}~\bibnamefont {Sun}}, \bibinfo
  {author} {\bibfnamefont {S.-N.}\ \bibnamefont {Sun}}, \bibinfo {author}
  {\bibfnamefont {S.}~\bibnamefont {Upadhyay}}, \bibinfo {author}
  {\bibfnamefont {L.~K.}\ \bibnamefont {Wagner}}, \bibinfo {author}
  {\bibfnamefont {X.}~\bibnamefont {Wang}}, \bibinfo {author} {\bibfnamefont
  {A.}~\bibnamefont {White}}, \bibinfo {author} {\bibfnamefont {J.~D.}\
  \bibnamefont {Whitfield}}, \bibinfo {author} {\bibfnamefont {M.~J.}\
  \bibnamefont {Williamson}}, \bibinfo {author} {\bibfnamefont
  {S.}~\bibnamefont {Wouters}}, \bibinfo {author} {\bibfnamefont
  {J.}~\bibnamefont {Yang}}, \bibinfo {author} {\bibfnamefont {J.~M.}\
  \bibnamefont {Yu}}, \bibinfo {author} {\bibfnamefont {T.}~\bibnamefont
  {Zhu}}, \bibinfo {author} {\bibfnamefont {T.~C.}\ \bibnamefont {Berkelbach}},
  \bibinfo {author} {\bibfnamefont {S.}~\bibnamefont {Sharma}}, \bibinfo
  {author} {\bibfnamefont {A.~Y.}\ \bibnamefont {Sokolov}}, \ and\ \bibinfo
  {author} {\bibfnamefont {G.~K.-L.}\ \bibnamefont {Chan}},\ }\bibfield
  {title} {\enquote {\bibinfo {title} {Recent developments in the pyscf program
  package},}\ }\href {\doibase 10.1063/5.0006074} {\bibfield  {journal}
  {\bibinfo  {journal} {The Journal of Chemical Physics}\ }\textbf {\bibinfo
  {volume} {153}},\ \bibinfo {pages} {024109} (\bibinfo {year} {2020})},\
  \Eprint
  {http://arxiv.org/abs/https://pubs.aip.org/aip/jcp/article-pdf/doi/10.1063/5.0006074/16722275/024109\_1\_online.pdf}
  {https://pubs.aip.org/aip/jcp/article-pdf/doi/10.1063/5.0006074/16722275/024109\_1\_online.pdf}
  \BibitemShut {NoStop}%
\bibitem [{\citenamefont {Smith}\ \emph {et~al.}(2018)\citenamefont {Smith},
  \citenamefont {Burns}, \citenamefont {Sirianni}, \citenamefont {Nascimento},
  \citenamefont {Kumar}, \citenamefont {James}, \citenamefont {Schriber},
  \citenamefont {Zhang}, \citenamefont {Zhang}, \citenamefont {Abbott},
  \citenamefont {Berquist}, \citenamefont {Lechner}, \citenamefont {Cunha},
  \citenamefont {Heide}, \citenamefont {Waldrop}, \citenamefont {Takeshita},
  \citenamefont {Alenaizan}, \citenamefont {Neuhauser}, \citenamefont {King},
  \citenamefont {Simmonett}, \citenamefont {Turney}, \citenamefont {Schaefer},
  \citenamefont {Evangelista}, \citenamefont {DePrince}, \citenamefont
  {Crawford}, \citenamefont {Patkowski},\ and\ \citenamefont
  {Sherrill}}]{Smith2018}%
  \BibitemOpen
  \bibfield  {author} {\bibinfo {author} {\bibfnamefont {D.~G.~A.}\
  \bibnamefont {Smith}}, \bibinfo {author} {\bibfnamefont {L.~A.}\ \bibnamefont
  {Burns}}, \bibinfo {author} {\bibfnamefont {D.~A.}\ \bibnamefont {Sirianni}},
  \bibinfo {author} {\bibfnamefont {D.~R.}\ \bibnamefont {Nascimento}},
  \bibinfo {author} {\bibfnamefont {A.}~\bibnamefont {Kumar}}, \bibinfo
  {author} {\bibfnamefont {A.~M.}\ \bibnamefont {James}}, \bibinfo {author}
  {\bibfnamefont {J.~B.}\ \bibnamefont {Schriber}}, \bibinfo {author}
  {\bibfnamefont {T.}~\bibnamefont {Zhang}}, \bibinfo {author} {\bibfnamefont
  {B.}~\bibnamefont {Zhang}}, \bibinfo {author} {\bibfnamefont {A.~S.}\
  \bibnamefont {Abbott}}, \bibinfo {author} {\bibfnamefont {E.~J.}\
  \bibnamefont {Berquist}}, \bibinfo {author} {\bibfnamefont {M.~H.}\
  \bibnamefont {Lechner}}, \bibinfo {author} {\bibfnamefont {L.~A.}\
  \bibnamefont {Cunha}}, \bibinfo {author} {\bibfnamefont {A.~G.}\ \bibnamefont
  {Heide}}, \bibinfo {author} {\bibfnamefont {J.~M.}\ \bibnamefont {Waldrop}},
  \bibinfo {author} {\bibfnamefont {T.~Y.}\ \bibnamefont {Takeshita}}, \bibinfo
  {author} {\bibfnamefont {A.}~\bibnamefont {Alenaizan}}, \bibinfo {author}
  {\bibfnamefont {D.}~\bibnamefont {Neuhauser}}, \bibinfo {author}
  {\bibfnamefont {R.~A.}\ \bibnamefont {King}}, \bibinfo {author}
  {\bibfnamefont {A.~C.}\ \bibnamefont {Simmonett}}, \bibinfo {author}
  {\bibfnamefont {J.~M.}\ \bibnamefont {Turney}}, \bibinfo {author}
  {\bibfnamefont {H.~F.}\ \bibnamefont {Schaefer}}, \bibinfo {author}
  {\bibfnamefont {F.~A.}\ \bibnamefont {Evangelista}}, \bibinfo {author}
  {\bibfnamefont {A.~E. I. I.~I.}\ \bibnamefont {DePrince}}, \bibinfo {author}
  {\bibfnamefont {T.~D.}\ \bibnamefont {Crawford}}, \bibinfo {author}
  {\bibfnamefont {K.}~\bibnamefont {Patkowski}}, \ and\ \bibinfo {author}
  {\bibfnamefont {C.~D.}\ \bibnamefont {Sherrill}},\ }\bibfield  {title}
  {\enquote {\bibinfo {title} {Psi4numpy: An interactive quantum chemistry
  programming environment for reference implementations and rapid
  development},}\ }\href {\doibase 10.1021/acs.jctc.8b00286} {\bibfield
  {journal} {\bibinfo  {journal} {J. Chem. Theory Comput.}\ }\textbf {\bibinfo
  {volume} {14}},\ \bibinfo {pages} {3504--3511} (\bibinfo {year}
  {2018})}\BibitemShut {NoStop}%
\bibitem [{\citenamefont {Boguslawski}\ \emph {et~al.}(2021)\citenamefont
  {Boguslawski}, \citenamefont {Leszczyk}, \citenamefont {Nowak}, \citenamefont
  {Brzęk}, \citenamefont {Żuchowski}, \citenamefont {Kędziera},\ and\
  \citenamefont {Tecmer}}]{Boguslawski2021}%
  \BibitemOpen
  \bibfield  {author} {\bibinfo {author} {\bibfnamefont {K.}~\bibnamefont
  {Boguslawski}}, \bibinfo {author} {\bibfnamefont {A.}~\bibnamefont
  {Leszczyk}}, \bibinfo {author} {\bibfnamefont {A.}~\bibnamefont {Nowak}},
  \bibinfo {author} {\bibfnamefont {F.}~\bibnamefont {Brzęk}}, \bibinfo
  {author} {\bibfnamefont {P.~S.}\ \bibnamefont {Żuchowski}}, \bibinfo
  {author} {\bibfnamefont {D.}~\bibnamefont {Kędziera}}, \ and\ \bibinfo
  {author} {\bibfnamefont {P.}~\bibnamefont {Tecmer}},\ }\bibfield  {title}
  {\enquote {\bibinfo {title} {Pythonic black-box electronic structure tool
  (pybest). an open-source python platform for electronic structure
  calculations at the interface between chemistry and physics},}\ }\href
  {https://www.sciencedirect.com/science/article/pii/S0010465521000643}
  {\bibfield  {journal} {\bibinfo  {journal} {Computer Physics Communications}\
  }\textbf {\bibinfo {volume} {264}},\ \bibinfo {pages} {107933} (\bibinfo
  {year} {2021})}\BibitemShut {NoStop}%
\bibitem [{\citenamefont {Hermann}\ \emph {et~al.}(2016)\citenamefont
  {Hermann}, \citenamefont {Pohl}, \citenamefont {Tremblay}, \citenamefont
  {Paulus}, \citenamefont {Hege},\ and\ \citenamefont {Schild}}]{Hermann2016}%
  \BibitemOpen
  \bibfield  {author} {\bibinfo {author} {\bibfnamefont {G.}~\bibnamefont
  {Hermann}}, \bibinfo {author} {\bibfnamefont {V.}~\bibnamefont {Pohl}},
  \bibinfo {author} {\bibfnamefont {J.~C.}\ \bibnamefont {Tremblay}}, \bibinfo
  {author} {\bibfnamefont {B.}~\bibnamefont {Paulus}}, \bibinfo {author}
  {\bibfnamefont {H.-C.}\ \bibnamefont {Hege}}, \ and\ \bibinfo {author}
  {\bibfnamefont {A.}~\bibnamefont {Schild}},\ }\bibfield  {title} {\enquote
  {\bibinfo {title} {Orbkit: A modular python toolbox for cross-platform
  postprocessing of quantum chemical wavefunction data},}\ }\href
  {https://doi.org/10.1002/jcc.24358} {\bibfield  {journal} {\bibinfo
  {journal} {J. Comput. Chem.}\ }\textbf {\bibinfo {volume} {37}},\ \bibinfo
  {pages} {1511--1520} (\bibinfo {year} {2016})}\BibitemShut {NoStop}%
\bibitem [{\citenamefont {Gao}\ \emph {et~al.}(2025)\citenamefont {Gao},
  \citenamefont {Fu}, \citenamefont {Jiao}, \citenamefont {Zhang},
  \citenamefont {Chen}, \citenamefont {Zhang}, \citenamefont {Wu},
  \citenamefont {Wan}, \citenamefont {Li}, \citenamefont {Hu},\ and\
  \citenamefont {Yang}}]{Gao2025}%
  \BibitemOpen
  \bibfield  {author} {\bibinfo {author} {\bibfnamefont {J.}~\bibnamefont
  {Gao}}, \bibinfo {author} {\bibfnamefont {L.}~\bibnamefont {Fu}}, \bibinfo
  {author} {\bibfnamefont {S.}~\bibnamefont {Jiao}}, \bibinfo {author}
  {\bibfnamefont {Z.}~\bibnamefont {Zhang}}, \bibinfo {author} {\bibfnamefont
  {S.}~\bibnamefont {Chen}}, \bibinfo {author} {\bibfnamefont {Z.}~\bibnamefont
  {Zhang}}, \bibinfo {author} {\bibfnamefont {W.}~\bibnamefont {Wu}}, \bibinfo
  {author} {\bibfnamefont {L.}~\bibnamefont {Wan}}, \bibinfo {author}
  {\bibfnamefont {J.}~\bibnamefont {Li}}, \bibinfo {author} {\bibfnamefont
  {W.}~\bibnamefont {Hu}}, \ and\ \bibinfo {author} {\bibfnamefont
  {J.}~\bibnamefont {Yang}},\ }\bibfield  {title} {\enquote {\bibinfo {title}
  {Pypwdft: A lightweight python software for single-node 10k atom plane-wave
  density functional theory calculations},}\ }\href {\doibase
  10.1021/acs.jctc.4c01605} {\bibfield  {journal} {\bibinfo  {journal} {J.
  Chem. Theory Comput.}\ }\textbf {\bibinfo {volume} {21}},\ \bibinfo {pages}
  {2353--2370} (\bibinfo {year} {2025})}\BibitemShut {NoStop}%
\bibitem [{\citenamefont {Evangelista}\ \emph {et~al.}(2024)\citenamefont
  {Evangelista}, \citenamefont {Li}, \citenamefont {Verma}, \citenamefont
  {Hannon}, \citenamefont {Schriber}, \citenamefont {Zhang}, \citenamefont
  {Cai}, \citenamefont {Wang}, \citenamefont {He}, \citenamefont {Stair},
  \citenamefont {Huang}, \citenamefont {Huang}, \citenamefont {Misiewicz},
  \citenamefont {Li}, \citenamefont {Marin}, \citenamefont {Zhao},\ and\
  \citenamefont {Burns}}]{Evangelista2024}%
  \BibitemOpen
  \bibfield  {author} {\bibinfo {author} {\bibfnamefont {F.~A.}\ \bibnamefont
  {Evangelista}}, \bibinfo {author} {\bibfnamefont {C.}~\bibnamefont {Li}},
  \bibinfo {author} {\bibfnamefont {P.}~\bibnamefont {Verma}}, \bibinfo
  {author} {\bibfnamefont {K.~P.}\ \bibnamefont {Hannon}}, \bibinfo {author}
  {\bibfnamefont {J.~B.}\ \bibnamefont {Schriber}}, \bibinfo {author}
  {\bibfnamefont {T.}~\bibnamefont {Zhang}}, \bibinfo {author} {\bibfnamefont
  {C.}~\bibnamefont {Cai}}, \bibinfo {author} {\bibfnamefont {S.}~\bibnamefont
  {Wang}}, \bibinfo {author} {\bibfnamefont {N.}~\bibnamefont {He}}, \bibinfo
  {author} {\bibfnamefont {N.~H.}\ \bibnamefont {Stair}}, \bibinfo {author}
  {\bibfnamefont {M.}~\bibnamefont {Huang}}, \bibinfo {author} {\bibfnamefont
  {R.}~\bibnamefont {Huang}}, \bibinfo {author} {\bibfnamefont {J.~P.}\
  \bibnamefont {Misiewicz}}, \bibinfo {author} {\bibfnamefont {S.}~\bibnamefont
  {Li}}, \bibinfo {author} {\bibfnamefont {K.}~\bibnamefont {Marin}}, \bibinfo
  {author} {\bibfnamefont {Z.}~\bibnamefont {Zhao}}, \ and\ \bibinfo {author}
  {\bibfnamefont {L.~A.}\ \bibnamefont {Burns}},\ }\bibfield  {title} {\enquote
  {\bibinfo {title} {Forte: A suite of advanced multireference quantum
  chemistry methods},}\ }\href {https://doi.org/10.1063/5.0216512} {\bibfield
  {journal} {\bibinfo  {journal} {J. Chem. Phys.}\ }\textbf {\bibinfo {volume}
  {161}},\ \bibinfo {pages} {062502} (\bibinfo {year} {2024})}\BibitemShut
  {NoStop}%
\bibitem [{\citenamefont {Chan}\ \emph {et~al.}(2024)\citenamefont {Chan},
  \citenamefont {Verstraelen}, \citenamefont {Tehrani}, \citenamefont {Richer},
  \citenamefont {Yang}, \citenamefont {Kim}, \citenamefont
  {Vöhringer-Martinez}, \citenamefont {Heidar-Zadeh},\ and\ \citenamefont
  {Ayers}}]{Chan2024}%
  \BibitemOpen
  \bibfield  {author} {\bibinfo {author} {\bibfnamefont {M.}~\bibnamefont
  {Chan}}, \bibinfo {author} {\bibfnamefont {T.}~\bibnamefont {Verstraelen}},
  \bibinfo {author} {\bibfnamefont {A.}~\bibnamefont {Tehrani}}, \bibinfo
  {author} {\bibfnamefont {M.}~\bibnamefont {Richer}}, \bibinfo {author}
  {\bibfnamefont {X.~D.}\ \bibnamefont {Yang}}, \bibinfo {author}
  {\bibfnamefont {T.~D.}\ \bibnamefont {Kim}}, \bibinfo {author} {\bibfnamefont
  {E.}~\bibnamefont {Vöhringer-Martinez}}, \bibinfo {author} {\bibfnamefont
  {F.}~\bibnamefont {Heidar-Zadeh}}, \ and\ \bibinfo {author} {\bibfnamefont
  {P.~W.}\ \bibnamefont {Ayers}},\ }\bibfield  {title} {\enquote {\bibinfo
  {title} {The tale of horton: Lessons learned in a decade of scientific
  software development},}\ }\href {https://doi.org/10.1063/5.0196638}
  {\bibfield  {journal} {\bibinfo  {journal} {J. Chem. Phys.}\ }\textbf
  {\bibinfo {volume} {160}},\ \bibinfo {pages} {162501} (\bibinfo {year}
  {2024})}\BibitemShut {NoStop}%
\bibitem [{\citenamefont {Posenitskiy}\ \emph {et~al.}(2023)\citenamefont
  {Posenitskiy}, \citenamefont {Chilkuri}, \citenamefont {Ammar}, \citenamefont
  {Hapka}, \citenamefont {Pernal}, \citenamefont {Shinde}, \citenamefont
  {Landinez~Borda}, \citenamefont {Filippi}, \citenamefont {Nakano},
  \citenamefont {Kohulák}, \citenamefont {Sorella}, \citenamefont
  {de~Oliveira~Castro}, \citenamefont {Jalby}, \citenamefont {Ríos},
  \citenamefont {Alavi},\ and\ \citenamefont {Scemama}}]{Posenitskiy2023}%
  \BibitemOpen
  \bibfield  {author} {\bibinfo {author} {\bibfnamefont {E.}~\bibnamefont
  {Posenitskiy}}, \bibinfo {author} {\bibfnamefont {V.~G.}\ \bibnamefont
  {Chilkuri}}, \bibinfo {author} {\bibfnamefont {A.}~\bibnamefont {Ammar}},
  \bibinfo {author} {\bibfnamefont {M.}~\bibnamefont {Hapka}}, \bibinfo
  {author} {\bibfnamefont {K.}~\bibnamefont {Pernal}}, \bibinfo {author}
  {\bibfnamefont {R.}~\bibnamefont {Shinde}}, \bibinfo {author} {\bibfnamefont
  {E.~J.}\ \bibnamefont {Landinez~Borda}}, \bibinfo {author} {\bibfnamefont
  {C.}~\bibnamefont {Filippi}}, \bibinfo {author} {\bibfnamefont
  {K.}~\bibnamefont {Nakano}}, \bibinfo {author} {\bibfnamefont
  {O.}~\bibnamefont {Kohulák}}, \bibinfo {author} {\bibfnamefont
  {S.}~\bibnamefont {Sorella}}, \bibinfo {author} {\bibfnamefont
  {P.}~\bibnamefont {de~Oliveira~Castro}}, \bibinfo {author} {\bibfnamefont
  {W.}~\bibnamefont {Jalby}}, \bibinfo {author} {\bibfnamefont {P.~L.}\
  \bibnamefont {Ríos}}, \bibinfo {author} {\bibfnamefont {A.}~\bibnamefont
  {Alavi}}, \ and\ \bibinfo {author} {\bibfnamefont {A.}~\bibnamefont
  {Scemama}},\ }\bibfield  {title} {\enquote {\bibinfo {title} {Trexio: A file
  format and library for quantum chemistry},}\ }\href
  {https://doi.org/10.1063/5.0148161} {\bibfield  {journal} {\bibinfo
  {journal} {J. Chem. Phys.}\ }\textbf {\bibinfo {volume} {158}},\ \bibinfo
  {pages} {174801} (\bibinfo {year} {2023})}\BibitemShut {NoStop}%
\bibitem [{\citenamefont {Zhang}\ and\ \citenamefont {Chan}(2022)}]{Zhang2022}%
  \BibitemOpen
  \bibfield  {author} {\bibinfo {author} {\bibfnamefont {X.}~\bibnamefont
  {Zhang}}\ and\ \bibinfo {author} {\bibfnamefont {G.~K.-L.}\ \bibnamefont
  {Chan}},\ }\bibfield  {title} {\enquote {\bibinfo {title} {Differentiable
  quantum chemistry with pyscf for molecules and materials at the mean-field
  level and beyond},}\ }\href {https://doi.org/10.1063/5.0118200} {\bibfield
  {journal} {\bibinfo  {journal} {J. Chem. Phys.}\ }\textbf {\bibinfo {volume}
  {157}},\ \bibinfo {pages} {204801} (\bibinfo {year} {2022})}\BibitemShut
  {NoStop}%
\bibitem [{\citenamefont {Kasim}, \citenamefont {Lehtola},\ and\ \citenamefont
  {Vinko}(2022)}]{Kasim2022}%
  \BibitemOpen
  \bibfield  {author} {\bibinfo {author} {\bibfnamefont {M.~F.}\ \bibnamefont
  {Kasim}}, \bibinfo {author} {\bibfnamefont {S.}~\bibnamefont {Lehtola}}, \
  and\ \bibinfo {author} {\bibfnamefont {S.~M.}\ \bibnamefont {Vinko}},\
  }\bibfield  {title} {\enquote {\bibinfo {title} {Dqc: A python program
  package for differentiable quantum chemistry},}\ }\href
  {https://doi.org/10.1063/5.0076202} {\bibfield  {journal} {\bibinfo
  {journal} {J. Chem. Phys.}\ }\textbf {\bibinfo {volume} {156}},\ \bibinfo
  {pages} {084801} (\bibinfo {year} {2022})}\BibitemShut {NoStop}%
\bibitem [{\citenamefont {Matteo}\ \emph {et~al.}(2023)\citenamefont {Matteo},
  \citenamefont {Izaac}, \citenamefont {Bromley}, \citenamefont {Hayes},
  \citenamefont {Lee}, \citenamefont {Schuld}, \citenamefont {Sz\'{a}va},
  \citenamefont {Roberts},\ and\ \citenamefont {Killoran}}]{Matteo2023}%
  \BibitemOpen
  \bibfield  {author} {\bibinfo {author} {\bibfnamefont {O.~D.}\ \bibnamefont
  {Matteo}}, \bibinfo {author} {\bibfnamefont {J.}~\bibnamefont {Izaac}},
  \bibinfo {author} {\bibfnamefont {T.~R.}\ \bibnamefont {Bromley}}, \bibinfo
  {author} {\bibfnamefont {A.}~\bibnamefont {Hayes}}, \bibinfo {author}
  {\bibfnamefont {C.}~\bibnamefont {Lee}}, \bibinfo {author} {\bibfnamefont
  {M.}~\bibnamefont {Schuld}}, \bibinfo {author} {\bibfnamefont
  {A.}~\bibnamefont {Sz\'{a}va}}, \bibinfo {author} {\bibfnamefont
  {C.}~\bibnamefont {Roberts}}, \ and\ \bibinfo {author} {\bibfnamefont
  {N.}~\bibnamefont {Killoran}},\ }\bibfield  {title} {\enquote {\bibinfo
  {title} {Quantum computing with differentiable quantum transforms},}\ }\href
  {\doibase 10.1145/3592622} {\bibfield  {journal} {\bibinfo  {journal} {ACM
  Transactions on Quantum Computing}\ }\textbf {\bibinfo {volume} {4}}
  (\bibinfo {year} {2023}),\ 10.1145/3592622}\BibitemShut {NoStop}%
\bibitem [{\citenamefont {Zhou}\ \emph {et~al.}(2020)\citenamefont {Zhou},
  \citenamefont {Nebgen}, \citenamefont {Lubbers}, \citenamefont {Malone},
  \citenamefont {Niklasson},\ and\ \citenamefont {Tretiak}}]{Zhou2020}%
  \BibitemOpen
  \bibfield  {author} {\bibinfo {author} {\bibfnamefont {G.}~\bibnamefont
  {Zhou}}, \bibinfo {author} {\bibfnamefont {B.}~\bibnamefont {Nebgen}},
  \bibinfo {author} {\bibfnamefont {N.}~\bibnamefont {Lubbers}}, \bibinfo
  {author} {\bibfnamefont {W.}~\bibnamefont {Malone}}, \bibinfo {author}
  {\bibfnamefont {A.~M.~N.}\ \bibnamefont {Niklasson}}, \ and\ \bibinfo
  {author} {\bibfnamefont {S.}~\bibnamefont {Tretiak}},\ }\bibfield  {title}
  {\enquote {\bibinfo {title} {Graphics processing unit-accelerated
  semiempirical born oppenheimer molecular dynamics using pytorch},}\ }\href
  {\doibase 10.1021/acs.jctc.0c00243} {\bibfield  {journal} {\bibinfo
  {journal} {J. Chem. Theory Comput.}\ }\textbf {\bibinfo {volume} {16}},\
  \bibinfo {pages} {4951--4962} (\bibinfo {year} {2020})}\BibitemShut {NoStop}%
\bibitem [{\citenamefont {Abbott}\ \emph {et~al.}(2021)\citenamefont {Abbott},
  \citenamefont {Abbott}, \citenamefont {Turney},\ and\ \citenamefont
  {Schaefer}}]{Abbott2021}%
  \BibitemOpen
  \bibfield  {author} {\bibinfo {author} {\bibfnamefont {A.~S.}\ \bibnamefont
  {Abbott}}, \bibinfo {author} {\bibfnamefont {B.~Z.}\ \bibnamefont {Abbott}},
  \bibinfo {author} {\bibfnamefont {J.~M.}\ \bibnamefont {Turney}}, \ and\
  \bibinfo {author} {\bibfnamefont {H.~F. I. I.~I.}\ \bibnamefont {Schaefer}},\
  }\bibfield  {title} {\enquote {\bibinfo {title} {Arbitrary-order derivatives
  of quantum chemical methods via automatic differentiation},}\ }\href
  {\doibase 10.1021/acs.jpclett.1c00607} {\bibfield  {journal} {\bibinfo
  {journal} {J. Phys. Chem. Lett.}\ }\textbf {\bibinfo {volume} {12}},\
  \bibinfo {pages} {3232--3239} (\bibinfo {year} {2021})}\BibitemShut {NoStop}%
\bibitem [{\citenamefont {Friede}\ \emph {et~al.}(2024)\citenamefont {Friede},
  \citenamefont {Hölzer}, \citenamefont {Ehlert},\ and\ \citenamefont
  {Grimme}}]{Friede2024}%
  \BibitemOpen
  \bibfield  {author} {\bibinfo {author} {\bibfnamefont {M.}~\bibnamefont
  {Friede}}, \bibinfo {author} {\bibfnamefont {C.}~\bibnamefont {Hölzer}},
  \bibinfo {author} {\bibfnamefont {S.}~\bibnamefont {Ehlert}}, \ and\ \bibinfo
  {author} {\bibfnamefont {S.}~\bibnamefont {Grimme}},\ }\bibfield  {title}
  {\enquote {\bibinfo {title} {dxtb--an efficient and fully differentiable
  framework for extended tight-binding},}\ }\href
  {https://doi.org/10.1063/5.0216715} {\bibfield  {journal} {\bibinfo
  {journal} {J. Chem. Phys.}\ }\textbf {\bibinfo {volume} {161}},\ \bibinfo
  {pages} {062501} (\bibinfo {year} {2024})}\BibitemShut {NoStop}%
\bibitem [{\citenamefont {Wu}\ \emph {et~al.}(2025)\citenamefont {Wu},
  \citenamefont {Sun}, \citenamefont {Pu}, \citenamefont {Zheng}, \citenamefont
  {Ma}, \citenamefont {Yan}, \citenamefont {Xia}, \citenamefont {Wu},
  \citenamefont {Huo}, \citenamefont {Li}, \citenamefont {Ren}, \citenamefont
  {Gong}, \citenamefont {Zhang},\ and\ \citenamefont {Gao}}]{Wu2025}%
  \BibitemOpen
  \bibfield  {author} {\bibinfo {author} {\bibfnamefont {X.}~\bibnamefont
  {Wu}}, \bibinfo {author} {\bibfnamefont {Q.}~\bibnamefont {Sun}}, \bibinfo
  {author} {\bibfnamefont {Z.}~\bibnamefont {Pu}}, \bibinfo {author}
  {\bibfnamefont {T.}~\bibnamefont {Zheng}}, \bibinfo {author} {\bibfnamefont
  {W.}~\bibnamefont {Ma}}, \bibinfo {author} {\bibfnamefont {W.}~\bibnamefont
  {Yan}}, \bibinfo {author} {\bibfnamefont {Y.}~\bibnamefont {Xia}}, \bibinfo
  {author} {\bibfnamefont {Z.}~\bibnamefont {Wu}}, \bibinfo {author}
  {\bibfnamefont {M.}~\bibnamefont {Huo}}, \bibinfo {author} {\bibfnamefont
  {X.}~\bibnamefont {Li}}, \bibinfo {author} {\bibfnamefont {W.}~\bibnamefont
  {Ren}}, \bibinfo {author} {\bibfnamefont {S.}~\bibnamefont {Gong}}, \bibinfo
  {author} {\bibfnamefont {Y.}~\bibnamefont {Zhang}}, \ and\ \bibinfo {author}
  {\bibfnamefont {W.}~\bibnamefont {Gao}},\ }\bibfield  {title} {\enquote
  {\bibinfo {title} {Enhancing gpu-acceleration in the python-based simulations
  of chemistry frameworks},}\ }\href {https://doi.org/10.1002/wcms.70008}
  {\bibfield  {journal} {\bibinfo  {journal} {WIREs Comput Mol Sci}\ }\textbf
  {\bibinfo {volume} {15}},\ \bibinfo {pages} {e70008} (\bibinfo {year}
  {2025})}\BibitemShut {NoStop}%
\bibitem [{\citenamefont {Kim}\ \emph {et~al.}(2024)\citenamefont {Kim},
  \citenamefont {Jeong}, \citenamefont {Weisburn}, \citenamefont {Alexiu},
  \citenamefont {Van~Voorhis}, \citenamefont {Rhee}, \citenamefont {Son},
  \citenamefont {Kim}, \citenamefont {Yim}, \citenamefont {Kim}, \citenamefont
  {Cho}, \citenamefont {Jang}, \citenamefont {Lee},\ and\ \citenamefont
  {Kim}}]{Kim2024}%
  \BibitemOpen
  \bibfield  {author} {\bibinfo {author} {\bibfnamefont {I.}~\bibnamefont
  {Kim}}, \bibinfo {author} {\bibfnamefont {D.}~\bibnamefont {Jeong}}, \bibinfo
  {author} {\bibfnamefont {L.~P.}\ \bibnamefont {Weisburn}}, \bibinfo {author}
  {\bibfnamefont {A.}~\bibnamefont {Alexiu}}, \bibinfo {author} {\bibfnamefont
  {T.}~\bibnamefont {Van~Voorhis}}, \bibinfo {author} {\bibfnamefont {Y.~M.}\
  \bibnamefont {Rhee}}, \bibinfo {author} {\bibfnamefont {W.-J.}\ \bibnamefont
  {Son}}, \bibinfo {author} {\bibfnamefont {H.-J.}\ \bibnamefont {Kim}},
  \bibinfo {author} {\bibfnamefont {J.}~\bibnamefont {Yim}}, \bibinfo {author}
  {\bibfnamefont {S.}~\bibnamefont {Kim}}, \bibinfo {author} {\bibfnamefont
  {Y.}~\bibnamefont {Cho}}, \bibinfo {author} {\bibfnamefont {I.}~\bibnamefont
  {Jang}}, \bibinfo {author} {\bibfnamefont {S.}~\bibnamefont {Lee}}, \ and\
  \bibinfo {author} {\bibfnamefont {D.~S.}\ \bibnamefont {Kim}},\ }\bibfield
  {title} {\enquote {\bibinfo {title} {Very-large-scale gpu-accelerated nuclear
  gradient of time-dependent density functional theory with tamm-dancoff
  approximation and range-separated hybrid functionals},}\ }\href {\doibase
  10.1021/acs.jctc.4c01003} {\bibfield  {journal} {\bibinfo  {journal} {J.
  Chem. Theory Comput.}\ }\textbf {\bibinfo {volume} {20}},\ \bibinfo {pages}
  {9018--9031} (\bibinfo {year} {2024})}\BibitemShut {NoStop}%
\bibitem [{\citenamefont {Tornai}\ \emph {et~al.}(2019)\citenamefont {Tornai},
  \citenamefont {Ladj\'anszki}, \citenamefont {R\'ak}, \citenamefont {Kis},\
  and\ \citenamefont {Cserey}}]{Tornai2019}%
  \BibitemOpen
  \bibfield  {author} {\bibinfo {author} {\bibfnamefont {G.}~\bibnamefont
  {Tornai}}, \bibinfo {author} {\bibfnamefont {I.}~\bibnamefont
  {Ladj\'anszki}}, \bibinfo {author} {\bibfnamefont {A.}~\bibnamefont {R\'ak}},
  \bibinfo {author} {\bibfnamefont {G.}~\bibnamefont {Kis}}, \ and\ \bibinfo
  {author} {\bibfnamefont {G.}~\bibnamefont {Cserey}},\ }\bibfield  {title}
  {\enquote {\bibinfo {title} {Calculation of quantum chemical two-electron
  integrals by applying compiler technology on gpu},}\ }\href {\doibase
  10.1021/acs.jctc.9b00560} {\bibfield  {journal} {\bibinfo  {journal} {J.
  Chem. Theory Comput.}\ }\textbf {\bibinfo {volume} {15}},\ \bibinfo {pages}
  {5319--5331} (\bibinfo {year} {2019})}\BibitemShut {NoStop}%
\bibitem [{\citenamefont {Li}\ \emph {et~al.}(2025)\citenamefont {Li},
  \citenamefont {Sun}, \citenamefont {Zhang},\ and\ \citenamefont
  {Chan}}]{Li2025a}%
  \BibitemOpen
  \bibfield  {author} {\bibinfo {author} {\bibfnamefont {R.}~\bibnamefont
  {Li}}, \bibinfo {author} {\bibfnamefont {Q.}~\bibnamefont {Sun}}, \bibinfo
  {author} {\bibfnamefont {X.}~\bibnamefont {Zhang}}, \ and\ \bibinfo {author}
  {\bibfnamefont {G.~K.-L.}\ \bibnamefont {Chan}},\ }\bibfield  {title}
  {\enquote {\bibinfo {title} {Introducing gpu acceleration into the
  python-based simulations of chemistry framework},}\ }\href {\doibase
  10.1021/acs.jpca.4c05876} {\bibfield  {journal} {\bibinfo  {journal} {J.
  Phys. Chem. A}\ }\textbf {\bibinfo {volume} {129}},\ \bibinfo {pages}
  {1459--1468} (\bibinfo {year} {2025})}\BibitemShut {NoStop}%
\bibitem [{\citenamefont {Vallejo}\ \emph {et~al.}(2023)\citenamefont
  {Vallejo}, \citenamefont {Jorge}, \citenamefont {Snowdon}, \citenamefont
  {Stocks}, \citenamefont {Kazemian}, \citenamefont {Yan~Yu}, \citenamefont
  {Seidl}, \citenamefont {Seeger}, \citenamefont {Alkan}, \citenamefont
  {Poole}, \citenamefont {Westheimer}, \citenamefont {Basha}, \citenamefont
  {De~La~Pierre}, \citenamefont {Rendell}, \citenamefont {Izgorodina},
  \citenamefont {Gordon},\ and\ \citenamefont {Barca}}]{Vallejo2023}%
  \BibitemOpen
  \bibfield  {author} {\bibinfo {author} {\bibfnamefont {G.}~\bibnamefont
  {Vallejo}}, \bibinfo {author} {\bibfnamefont {L.}~\bibnamefont {Jorge}},
  \bibinfo {author} {\bibfnamefont {C.}~\bibnamefont {Snowdon}}, \bibinfo
  {author} {\bibfnamefont {R.}~\bibnamefont {Stocks}}, \bibinfo {author}
  {\bibfnamefont {F.}~\bibnamefont {Kazemian}}, \bibinfo {author}
  {\bibfnamefont {F.~C.}\ \bibnamefont {Yan~Yu}}, \bibinfo {author}
  {\bibfnamefont {C.}~\bibnamefont {Seidl}}, \bibinfo {author} {\bibfnamefont
  {Z.}~\bibnamefont {Seeger}}, \bibinfo {author} {\bibfnamefont
  {M.}~\bibnamefont {Alkan}}, \bibinfo {author} {\bibfnamefont
  {D.}~\bibnamefont {Poole}}, \bibinfo {author} {\bibfnamefont {B.~M.}\
  \bibnamefont {Westheimer}}, \bibinfo {author} {\bibfnamefont
  {M.}~\bibnamefont {Basha}}, \bibinfo {author} {\bibfnamefont
  {M.}~\bibnamefont {De~La~Pierre}}, \bibinfo {author} {\bibfnamefont
  {A.}~\bibnamefont {Rendell}}, \bibinfo {author} {\bibfnamefont {E.~I.}\
  \bibnamefont {Izgorodina}}, \bibinfo {author} {\bibfnamefont {M.~S.}\
  \bibnamefont {Gordon}}, \ and\ \bibinfo {author} {\bibfnamefont {G.~M.~J.}\
  \bibnamefont {Barca}},\ }\bibfield  {title} {\enquote {\bibinfo {title}
  {Toward an extreme-scale electronic structure system},}\ }\href
  {https://doi.org/10.1063/5.0156399} {\bibfield  {journal} {\bibinfo
  {journal} {J. Chem. Phys.}\ }\textbf {\bibinfo {volume} {159}},\ \bibinfo
  {pages} {044112} (\bibinfo {year} {2023})}\BibitemShut {NoStop}%
\bibitem [{\citenamefont {Seritan}\ \emph {et~al.}(2021)\citenamefont
  {Seritan}, \citenamefont {Bannwarth}, \citenamefont {Fales}, \citenamefont
  {Hohenstein}, \citenamefont {Isborn}, \citenamefont {Kokkila-Schumacher},
  \citenamefont {Li}, \citenamefont {Liu}, \citenamefont {Luehr}, \citenamefont
  {Snyder~Jr.}, \citenamefont {Song}, \citenamefont {Titov}, \citenamefont
  {Ufimtsev}, \citenamefont {Wang},\ and\ \citenamefont
  {Martínez}}]{Seritan2021}%
  \BibitemOpen
  \bibfield  {author} {\bibinfo {author} {\bibfnamefont {S.}~\bibnamefont
  {Seritan}}, \bibinfo {author} {\bibfnamefont {C.}~\bibnamefont {Bannwarth}},
  \bibinfo {author} {\bibfnamefont {B.~S.}\ \bibnamefont {Fales}}, \bibinfo
  {author} {\bibfnamefont {E.~G.}\ \bibnamefont {Hohenstein}}, \bibinfo
  {author} {\bibfnamefont {C.~M.}\ \bibnamefont {Isborn}}, \bibinfo {author}
  {\bibfnamefont {S.~I.~L.}\ \bibnamefont {Kokkila-Schumacher}}, \bibinfo
  {author} {\bibfnamefont {X.}~\bibnamefont {Li}}, \bibinfo {author}
  {\bibfnamefont {F.}~\bibnamefont {Liu}}, \bibinfo {author} {\bibfnamefont
  {N.}~\bibnamefont {Luehr}}, \bibinfo {author} {\bibfnamefont {J.~W.}\
  \bibnamefont {Snyder~Jr.}}, \bibinfo {author} {\bibfnamefont
  {C.}~\bibnamefont {Song}}, \bibinfo {author} {\bibfnamefont {A.~V.}\
  \bibnamefont {Titov}}, \bibinfo {author} {\bibfnamefont {I.~S.}\ \bibnamefont
  {Ufimtsev}}, \bibinfo {author} {\bibfnamefont {L.-P.}\ \bibnamefont {Wang}},
  \ and\ \bibinfo {author} {\bibfnamefont {T.~J.}\ \bibnamefont {Martínez}},\
  }\bibfield  {title} {\enquote {\bibinfo {title} {Terachem: A graphical
  processing unit-accelerated electronic structure package for large-scale ab
  initio molecular dynamics},}\ }\href {https://doi.org/10.1002/wcms.1494}
  {\bibfield  {journal} {\bibinfo  {journal} {WIREs Comput Mol Sci}\ }\textbf
  {\bibinfo {volume} {11}},\ \bibinfo {pages} {e1494} (\bibinfo {year}
  {2021})}\BibitemShut {NoStop}%
\bibitem [{\citenamefont {Poole}, \citenamefont {Galvez~Vallejo},\ and\
  \citenamefont {Gordon}(2020)}]{Poole2020}%
  \BibitemOpen
  \bibfield  {author} {\bibinfo {author} {\bibfnamefont {D.}~\bibnamefont
  {Poole}}, \bibinfo {author} {\bibfnamefont {J.~L.}\ \bibnamefont
  {Galvez~Vallejo}}, \ and\ \bibinfo {author} {\bibfnamefont {M.~S.}\
  \bibnamefont {Gordon}},\ }\bibfield  {title} {\enquote {\bibinfo {title} {A
  new kid on the block: Application of julia to hartree-fock calculations},}\
  }\href {\doibase 10.1021/acs.jctc.0c00337} {\bibfield  {journal} {\bibinfo
  {journal} {J. Chem. Theory Comput.}\ }\textbf {\bibinfo {volume} {16}},\
  \bibinfo {pages} {5006--5013} (\bibinfo {year} {2020})}\BibitemShut {NoStop}%
\bibitem [{\citenamefont {Aroeira}\ \emph {et~al.}(2022)\citenamefont
  {Aroeira}, \citenamefont {Davis}, \citenamefont {Turney},\ and\ \citenamefont
  {Schaefer}}]{Aroeira2022}%
  \BibitemOpen
  \bibfield  {author} {\bibinfo {author} {\bibfnamefont {G.~J.~R.}\
  \bibnamefont {Aroeira}}, \bibinfo {author} {\bibfnamefont {M.~M.}\
  \bibnamefont {Davis}}, \bibinfo {author} {\bibfnamefont {J.~M.}\ \bibnamefont
  {Turney}}, \ and\ \bibinfo {author} {\bibfnamefont {H.~F. I. I.~I.}\
  \bibnamefont {Schaefer}},\ }\bibfield  {title} {\enquote {\bibinfo {title}
  {Fermi.jl: A modern design for quantum chemistry},}\ }\href {\doibase
  10.1021/acs.jctc.1c00719} {\bibfield  {journal} {\bibinfo  {journal} {J.
  Chem. Theory Comput.}\ }\textbf {\bibinfo {volume} {18}},\ \bibinfo {pages}
  {677--686} (\bibinfo {year} {2022})}\BibitemShut {NoStop}%
\bibitem [{\citenamefont {Wang}\ and\ \citenamefont
  {Whitfield}(2023)}]{Wang2023}%
  \BibitemOpen
  \bibfield  {author} {\bibinfo {author} {\bibfnamefont {W.}~\bibnamefont
  {Wang}}\ and\ \bibinfo {author} {\bibfnamefont {J.~D.}\ \bibnamefont
  {Whitfield}},\ }\bibfield  {title} {\enquote {\bibinfo {title} {Basis set
  generation and optimization in the nisq era with quiqbox.jl},}\ }\href
  {\doibase 10.1021/acs.jctc.3c00011} {\bibfield  {journal} {\bibinfo
  {journal} {J. Chem. Theory Comput.}\ }\textbf {\bibinfo {volume} {19}},\
  \bibinfo {pages} {8032--8052} (\bibinfo {year} {2023})}\BibitemShut {NoStop}%
\bibitem [{\citenamefont {Wang}\ \emph {et~al.}(2024)\citenamefont {Wang},
  \citenamefont {Lin}, \citenamefont {Ouyang}, \citenamefont {Jiang},
  \citenamefont {Zhang},\ and\ \citenamefont {Xu}}]{Wang2024}%
  \BibitemOpen
  \bibfield  {author} {\bibinfo {author} {\bibfnamefont {Y.}~\bibnamefont
  {Wang}}, \bibinfo {author} {\bibfnamefont {Z.}~\bibnamefont {Lin}}, \bibinfo
  {author} {\bibfnamefont {R.}~\bibnamefont {Ouyang}}, \bibinfo {author}
  {\bibfnamefont {B.}~\bibnamefont {Jiang}}, \bibinfo {author} {\bibfnamefont
  {I.~Y.}\ \bibnamefont {Zhang}}, \ and\ \bibinfo {author} {\bibfnamefont
  {X.}~\bibnamefont {Xu}},\ }\bibfield  {title} {\enquote {\bibinfo {title}
  {Toward efficient and unified treatment of static and dynamic correlations in
  generalized kohn-sham density functional theory},}\ }\href {\doibase
  10.1021/jacsau.4c00488} {\bibfield  {journal} {\bibinfo  {journal} {JACS Au}\
  }\textbf {\bibinfo {volume} {4}},\ \bibinfo {pages} {3205--3216} (\bibinfo
  {year} {2024})}\BibitemShut {NoStop}%
\bibitem [{\citenamefont {Neese}\ \emph {et~al.}(2020)\citenamefont {Neese},
  \citenamefont {Wennmohs}, \citenamefont {Becker},\ and\ \citenamefont
  {Riplinger}}]{Neese2020}%
  \BibitemOpen
  \bibfield  {author} {\bibinfo {author} {\bibfnamefont {F.}~\bibnamefont
  {Neese}}, \bibinfo {author} {\bibfnamefont {F.}~\bibnamefont {Wennmohs}},
  \bibinfo {author} {\bibfnamefont {U.}~\bibnamefont {Becker}}, \ and\ \bibinfo
  {author} {\bibfnamefont {C.}~\bibnamefont {Riplinger}},\ }\bibfield  {title}
  {\enquote {\bibinfo {title} {The orca quantum chemistry program package},}\
  }\href {https://doi.org/10.1063/5.0004608} {\bibfield  {journal} {\bibinfo
  {journal} {J. Chem. Phys.}\ }\textbf {\bibinfo {volume} {152}},\ \bibinfo
  {pages} {224108} (\bibinfo {year} {2020})}\BibitemShut {NoStop}%
\bibitem [{\citenamefont {Peng}\ \emph {et~al.}(2020)\citenamefont {Peng},
  \citenamefont {Lewis}, \citenamefont {Wang}, \citenamefont {Clement},
  \citenamefont {Pierce}, \citenamefont {Rishi}, \citenamefont {Pavošević},
  \citenamefont {Slattery}, \citenamefont {Zhang}, \citenamefont {Teke},
  \citenamefont {Kumar}, \citenamefont {Masteran}, \citenamefont {Asadchev},
  \citenamefont {Calvin},\ and\ \citenamefont {Valeev}}]{Peng2020}%
  \BibitemOpen
  \bibfield  {author} {\bibinfo {author} {\bibfnamefont {C.}~\bibnamefont
  {Peng}}, \bibinfo {author} {\bibfnamefont {C.~A.}\ \bibnamefont {Lewis}},
  \bibinfo {author} {\bibfnamefont {X.}~\bibnamefont {Wang}}, \bibinfo {author}
  {\bibfnamefont {M.~C.}\ \bibnamefont {Clement}}, \bibinfo {author}
  {\bibfnamefont {K.}~\bibnamefont {Pierce}}, \bibinfo {author} {\bibfnamefont
  {V.}~\bibnamefont {Rishi}}, \bibinfo {author} {\bibfnamefont
  {F.}~\bibnamefont {Pavošević}}, \bibinfo {author} {\bibfnamefont
  {S.}~\bibnamefont {Slattery}}, \bibinfo {author} {\bibfnamefont
  {J.}~\bibnamefont {Zhang}}, \bibinfo {author} {\bibfnamefont
  {N.}~\bibnamefont {Teke}}, \bibinfo {author} {\bibfnamefont {A.}~\bibnamefont
  {Kumar}}, \bibinfo {author} {\bibfnamefont {C.}~\bibnamefont {Masteran}},
  \bibinfo {author} {\bibfnamefont {A.}~\bibnamefont {Asadchev}}, \bibinfo
  {author} {\bibfnamefont {J.~A.}\ \bibnamefont {Calvin}}, \ and\ \bibinfo
  {author} {\bibfnamefont {E.~F.}\ \bibnamefont {Valeev}},\ }\bibfield  {title}
  {\enquote {\bibinfo {title} {Massively parallel quantum chemistry: A
  high-performance research platform for electronic structure},}\ }\href
  {https://doi.org/10.1063/5.0005889} {\bibfield  {journal} {\bibinfo
  {journal} {J. Chem. Phys.}\ }\textbf {\bibinfo {volume} {153}},\ \bibinfo
  {pages} {044120} (\bibinfo {year} {2020})}\BibitemShut {NoStop}%
\bibitem [{\citenamefont {Smith}\ \emph {et~al.}(2020)\citenamefont {Smith},
  \citenamefont {Burns}, \citenamefont {Simmonett}, \citenamefont {Parrish},
  \citenamefont {Schieber}, \citenamefont {Galvelis}, \citenamefont {Kraus},
  \citenamefont {Kruse}, \citenamefont {Di~Remigio}, \citenamefont {Alenaizan},
  \citenamefont {James}, \citenamefont {Lehtola}, \citenamefont {Misiewicz},
  \citenamefont {Scheurer}, \citenamefont {Shaw}, \citenamefont {Schriber},
  \citenamefont {Xie}, \citenamefont {Glick}, \citenamefont {Sirianni},
  \citenamefont {O’Brien}, \citenamefont {Waldrop}, \citenamefont {Kumar},
  \citenamefont {Hohenstein}, \citenamefont {Pritchard}, \citenamefont
  {Brooks}, \citenamefont {Schaefer}, \citenamefont {Sokolov}, \citenamefont
  {Patkowski}, \citenamefont {DePrince}, \citenamefont {Bozkaya}, \citenamefont
  {King}, \citenamefont {Evangelista}, \citenamefont {Turney}, \citenamefont
  {Crawford},\ and\ \citenamefont {Sherrill}}]{Smith2020}%
  \BibitemOpen
  \bibfield  {author} {\bibinfo {author} {\bibfnamefont {D.~G.~A.}\
  \bibnamefont {Smith}}, \bibinfo {author} {\bibfnamefont {L.~A.}\ \bibnamefont
  {Burns}}, \bibinfo {author} {\bibfnamefont {A.~C.}\ \bibnamefont
  {Simmonett}}, \bibinfo {author} {\bibfnamefont {R.~M.}\ \bibnamefont
  {Parrish}}, \bibinfo {author} {\bibfnamefont {M.~C.}\ \bibnamefont
  {Schieber}}, \bibinfo {author} {\bibfnamefont {R.}~\bibnamefont {Galvelis}},
  \bibinfo {author} {\bibfnamefont {P.}~\bibnamefont {Kraus}}, \bibinfo
  {author} {\bibfnamefont {H.}~\bibnamefont {Kruse}}, \bibinfo {author}
  {\bibfnamefont {R.}~\bibnamefont {Di~Remigio}}, \bibinfo {author}
  {\bibfnamefont {A.}~\bibnamefont {Alenaizan}}, \bibinfo {author}
  {\bibfnamefont {A.~M.}\ \bibnamefont {James}}, \bibinfo {author}
  {\bibfnamefont {S.}~\bibnamefont {Lehtola}}, \bibinfo {author} {\bibfnamefont
  {J.~P.}\ \bibnamefont {Misiewicz}}, \bibinfo {author} {\bibfnamefont
  {M.}~\bibnamefont {Scheurer}}, \bibinfo {author} {\bibfnamefont {R.~A.}\
  \bibnamefont {Shaw}}, \bibinfo {author} {\bibfnamefont {J.~B.}\ \bibnamefont
  {Schriber}}, \bibinfo {author} {\bibfnamefont {Y.}~\bibnamefont {Xie}},
  \bibinfo {author} {\bibfnamefont {Z.~L.}\ \bibnamefont {Glick}}, \bibinfo
  {author} {\bibfnamefont {D.~A.}\ \bibnamefont {Sirianni}}, \bibinfo {author}
  {\bibfnamefont {J.~S.}\ \bibnamefont {O’Brien}}, \bibinfo {author}
  {\bibfnamefont {J.~M.}\ \bibnamefont {Waldrop}}, \bibinfo {author}
  {\bibfnamefont {A.}~\bibnamefont {Kumar}}, \bibinfo {author} {\bibfnamefont
  {E.~G.}\ \bibnamefont {Hohenstein}}, \bibinfo {author} {\bibfnamefont
  {B.~P.}\ \bibnamefont {Pritchard}}, \bibinfo {author} {\bibfnamefont {B.~R.}\
  \bibnamefont {Brooks}}, \bibinfo {author} {\bibfnamefont {I.~I.~I.}\
  \bibnamefont {Schaefer}, \bibfnamefont {Henry~F.}}, \bibinfo {author}
  {\bibfnamefont {A.~Y.}\ \bibnamefont {Sokolov}}, \bibinfo {author}
  {\bibfnamefont {K.}~\bibnamefont {Patkowski}}, \bibinfo {author}
  {\bibfnamefont {I.~I.~I.}\ \bibnamefont {DePrince}, \bibfnamefont
  {A.~Eugene}}, \bibinfo {author} {\bibfnamefont {U.}~\bibnamefont {Bozkaya}},
  \bibinfo {author} {\bibfnamefont {R.~A.}\ \bibnamefont {King}}, \bibinfo
  {author} {\bibfnamefont {F.~A.}\ \bibnamefont {Evangelista}}, \bibinfo
  {author} {\bibfnamefont {J.~M.}\ \bibnamefont {Turney}}, \bibinfo {author}
  {\bibfnamefont {T.~D.}\ \bibnamefont {Crawford}}, \ and\ \bibinfo {author}
  {\bibfnamefont {C.~D.}\ \bibnamefont {Sherrill}},\ }\bibfield  {title}
  {\enquote {\bibinfo {title} {Psi4 1.4: Open-source software for
  high-throughput quantum chemistry},}\ }\href
  {https://doi.org/10.1063/5.0006002} {\bibfield  {journal} {\bibinfo
  {journal} {J. Chem. Phys.}\ }\textbf {\bibinfo {volume} {152}},\ \bibinfo
  {pages} {184108} (\bibinfo {year} {2020})}\BibitemShut {NoStop}%
\bibitem [{\citenamefont {Shiozaki}(2018)}]{Shiozaki2018}%
  \BibitemOpen
  \bibfield  {author} {\bibinfo {author} {\bibfnamefont {T.}~\bibnamefont
  {Shiozaki}},\ }\bibfield  {title} {\enquote {\bibinfo {title} {Bagel:
  Brilliantly advanced general electronic-structure library},}\ }\href
  {https://doi.org/10.1002/wcms.1331} {\bibfield  {journal} {\bibinfo
  {journal} {WIREs Comput Mol Sci}\ }\textbf {\bibinfo {volume} {8}},\ \bibinfo
  {pages} {e1331} (\bibinfo {year} {2018})}\BibitemShut {NoStop}%
\bibitem [{\citenamefont {Williams-Young}\ \emph {et~al.}(2020)\citenamefont
  {Williams-Young}, \citenamefont {Petrone}, \citenamefont {Sun}, \citenamefont
  {Stetina}, \citenamefont {Lestrange}, \citenamefont {Hoyer}, \citenamefont
  {Nascimento}, \citenamefont {Koulias}, \citenamefont {Wildman}, \citenamefont
  {Kasper}, \citenamefont {Goings}, \citenamefont {Ding}, \citenamefont
  {DePrince~III}, \citenamefont {Valeev},\ and\ \citenamefont
  {Li}}]{Williams-Young2020}%
  \BibitemOpen
  \bibfield  {author} {\bibinfo {author} {\bibfnamefont {D.~B.}\ \bibnamefont
  {Williams-Young}}, \bibinfo {author} {\bibfnamefont {A.}~\bibnamefont
  {Petrone}}, \bibinfo {author} {\bibfnamefont {S.}~\bibnamefont {Sun}},
  \bibinfo {author} {\bibfnamefont {T.~F.}\ \bibnamefont {Stetina}}, \bibinfo
  {author} {\bibfnamefont {P.}~\bibnamefont {Lestrange}}, \bibinfo {author}
  {\bibfnamefont {C.~E.}\ \bibnamefont {Hoyer}}, \bibinfo {author}
  {\bibfnamefont {D.~R.}\ \bibnamefont {Nascimento}}, \bibinfo {author}
  {\bibfnamefont {L.}~\bibnamefont {Koulias}}, \bibinfo {author} {\bibfnamefont
  {A.}~\bibnamefont {Wildman}}, \bibinfo {author} {\bibfnamefont
  {J.}~\bibnamefont {Kasper}}, \bibinfo {author} {\bibfnamefont {J.~J.}\
  \bibnamefont {Goings}}, \bibinfo {author} {\bibfnamefont {F.}~\bibnamefont
  {Ding}}, \bibinfo {author} {\bibfnamefont {A.~E.}\ \bibnamefont
  {DePrince~III}}, \bibinfo {author} {\bibfnamefont {E.~F.}\ \bibnamefont
  {Valeev}}, \ and\ \bibinfo {author} {\bibfnamefont {X.}~\bibnamefont {Li}},\
  }\bibfield  {title} {\enquote {\bibinfo {title} {The chronus quantum software
  package},}\ }\href {https://doi.org/10.1002/wcms.1436} {\bibfield  {journal}
  {\bibinfo  {journal} {WIREs Comput Mol Sci}\ }\textbf {\bibinfo {volume}
  {10}},\ \bibinfo {pages} {e1436} (\bibinfo {year} {2020})}\BibitemShut
  {NoStop}%
\bibitem [{\citenamefont {Dral}(2020)}]{Dral2020}%
  \BibitemOpen
  \bibfield  {author} {\bibinfo {author} {\bibfnamefont {P.~O.}\ \bibnamefont
  {Dral}},\ }\bibfield  {title} {\enquote {\bibinfo {title} {Quantum chemistry
  in the age of machine learning},}\ }\href {\doibase
  10.1021/acs.jpclett.9b03664} {\bibfield  {journal} {\bibinfo  {journal} {J.
  Phys. Chem. Lett.}\ }\textbf {\bibinfo {volume} {11}},\ \bibinfo {pages}
  {2336--2347} (\bibinfo {year} {2020})}\BibitemShut {NoStop}%
\bibitem [{\citenamefont {Gong}\ \emph {et~al.}(2025)\citenamefont {Gong},
  \citenamefont {Zhang}, \citenamefont {Mu}, \citenamefont {Pu}, \citenamefont
  {Wang}, \citenamefont {Han}, \citenamefont {Yu}, \citenamefont {Chen},
  \citenamefont {Zheng}, \citenamefont {Wang}, \citenamefont {Chen},
  \citenamefont {Yang}, \citenamefont {Wu}, \citenamefont {Shi}, \citenamefont
  {Gao}, \citenamefont {Yan},\ and\ \citenamefont {Xiang}}]{Gong2025}%
  \BibitemOpen
  \bibfield  {author} {\bibinfo {author} {\bibfnamefont {S.}~\bibnamefont
  {Gong}}, \bibinfo {author} {\bibfnamefont {Y.}~\bibnamefont {Zhang}},
  \bibinfo {author} {\bibfnamefont {Z.}~\bibnamefont {Mu}}, \bibinfo {author}
  {\bibfnamefont {Z.}~\bibnamefont {Pu}}, \bibinfo {author} {\bibfnamefont
  {H.}~\bibnamefont {Wang}}, \bibinfo {author} {\bibfnamefont {X.}~\bibnamefont
  {Han}}, \bibinfo {author} {\bibfnamefont {Z.}~\bibnamefont {Yu}}, \bibinfo
  {author} {\bibfnamefont {M.}~\bibnamefont {Chen}}, \bibinfo {author}
  {\bibfnamefont {T.}~\bibnamefont {Zheng}}, \bibinfo {author} {\bibfnamefont
  {Z.}~\bibnamefont {Wang}}, \bibinfo {author} {\bibfnamefont {L.}~\bibnamefont
  {Chen}}, \bibinfo {author} {\bibfnamefont {Z.}~\bibnamefont {Yang}}, \bibinfo
  {author} {\bibfnamefont {X.}~\bibnamefont {Wu}}, \bibinfo {author}
  {\bibfnamefont {S.}~\bibnamefont {Shi}}, \bibinfo {author} {\bibfnamefont
  {W.}~\bibnamefont {Gao}}, \bibinfo {author} {\bibfnamefont {W.}~\bibnamefont
  {Yan}}, \ and\ \bibinfo {author} {\bibfnamefont {L.}~\bibnamefont {Xiang}},\
  }\bibfield  {title} {\enquote {\bibinfo {title} {A predictive machine
  learning force-field framework for liquid electrolyte development},}\ }\href
  {https://doi.org/10.1038/s42256-025-01009-7} {\bibfield  {journal} {\bibinfo
  {journal} {Nature Machine Intelligence}\ }\textbf {\bibinfo {volume} {7}},\
  \bibinfo {pages} {543--552} (\bibinfo {year} {2025})}\BibitemShut {NoStop}%
\bibitem [{\citenamefont {Chen}\ \emph {et~al.}(2024)\citenamefont {Chen},
  \citenamefont {Yan}, \citenamefont {Wang}, \citenamefont {Wu},\ and\
  \citenamefont {Xu}}]{Chen2024}%
  \BibitemOpen
  \bibfield  {author} {\bibinfo {author} {\bibfnamefont {Y.}~\bibnamefont
  {Chen}}, \bibinfo {author} {\bibfnamefont {W.}~\bibnamefont {Yan}}, \bibinfo
  {author} {\bibfnamefont {Z.}~\bibnamefont {Wang}}, \bibinfo {author}
  {\bibfnamefont {J.}~\bibnamefont {Wu}}, \ and\ \bibinfo {author}
  {\bibfnamefont {X.}~\bibnamefont {Xu}},\ }\bibfield  {title} {\enquote
  {\bibinfo {title} {Constructing accurate and efficient general-purpose
  atomistic machine learning model with transferable accuracy for quantum
  chemistry},}\ }\href {\doibase 10.1021/acs.jctc.4c01151} {\bibfield
  {journal} {\bibinfo  {journal} {J. Chem. Theory Comput.}\ }\textbf {\bibinfo
  {volume} {20}},\ \bibinfo {pages} {9500--9511} (\bibinfo {year}
  {2024})}\BibitemShut {NoStop}%
\bibitem [{\citenamefont {Khan}\ \emph {et~al.}(2025)\citenamefont {Khan},
  \citenamefont {Price}, \citenamefont {Huang}, \citenamefont {Ach},\ and\
  \citenamefont {von Lilienfeld}}]{Khan2025}%
  \BibitemOpen
  \bibfield  {author} {\bibinfo {author} {\bibfnamefont {D.}~\bibnamefont
  {Khan}}, \bibinfo {author} {\bibfnamefont {A.~J.~A.}\ \bibnamefont {Price}},
  \bibinfo {author} {\bibfnamefont {B.}~\bibnamefont {Huang}}, \bibinfo
  {author} {\bibfnamefont {M.~L.}\ \bibnamefont {Ach}}, \ and\ \bibinfo
  {author} {\bibfnamefont {O.~A.}\ \bibnamefont {von Lilienfeld}},\ }\bibfield
  {title} {\enquote {\bibinfo {title} {Adapting hybrid density functionals with
  machine learning},}\ }\href {\doibase 10.1126/sciadv.adt7769} {\bibfield
  {journal} {\bibinfo  {journal} {Science Advances}\ }\textbf {\bibinfo
  {volume} {11}},\ \bibinfo {pages} {eadt7769} (\bibinfo {year}
  {2025})}\BibitemShut {NoStop}%
\bibitem [{\citenamefont {Wu}\ \emph {et~al.}(2023)\citenamefont {Wu},
  \citenamefont {Pun}, \citenamefont {Zheng},\ and\ \citenamefont
  {Chen}}]{Wu2023}%
  \BibitemOpen
  \bibfield  {author} {\bibinfo {author} {\bibfnamefont {J.}~\bibnamefont
  {Wu}}, \bibinfo {author} {\bibfnamefont {S.-M.}\ \bibnamefont {Pun}},
  \bibinfo {author} {\bibfnamefont {X.}~\bibnamefont {Zheng}}, \ and\ \bibinfo
  {author} {\bibfnamefont {G.}~\bibnamefont {Chen}},\ }\bibfield  {title}
  {\enquote {\bibinfo {title} {Construct exchange-correlation functional via
  machine learning},}\ }\href {https://doi.org/10.1063/5.0150587} {\bibfield
  {journal} {\bibinfo  {journal} {J. Chem. Phys.}\ }\textbf {\bibinfo {volume}
  {159}},\ \bibinfo {pages} {090901} (\bibinfo {year} {2023})}\BibitemShut
  {NoStop}%
\bibitem [{\citenamefont {Kirkpatrick}\ \emph {et~al.}(2021)\citenamefont
  {Kirkpatrick}, \citenamefont {McMorrow}, \citenamefont {Turban},
  \citenamefont {Gaunt}, \citenamefont {Spencer}, \citenamefont {Matthews},
  \citenamefont {Obika}, \citenamefont {Thiry}, \citenamefont {Fortunato},
  \citenamefont {Pfau}, \citenamefont {Castellanos}, \citenamefont {Petersen},
  \citenamefont {Nelson}, \citenamefont {Kohli}, \citenamefont {Mori-Sánchez},
  \citenamefont {Hassabis},\ and\ \citenamefont {Cohen}}]{Kirkpatrick2021}%
  \BibitemOpen
  \bibfield  {author} {\bibinfo {author} {\bibfnamefont {J.}~\bibnamefont
  {Kirkpatrick}}, \bibinfo {author} {\bibfnamefont {B.}~\bibnamefont
  {McMorrow}}, \bibinfo {author} {\bibfnamefont {D.~H.~P.}\ \bibnamefont
  {Turban}}, \bibinfo {author} {\bibfnamefont {A.~L.}\ \bibnamefont {Gaunt}},
  \bibinfo {author} {\bibfnamefont {J.~S.}\ \bibnamefont {Spencer}}, \bibinfo
  {author} {\bibfnamefont {A.~G. D.~G.}\ \bibnamefont {Matthews}}, \bibinfo
  {author} {\bibfnamefont {A.}~\bibnamefont {Obika}}, \bibinfo {author}
  {\bibfnamefont {L.}~\bibnamefont {Thiry}}, \bibinfo {author} {\bibfnamefont
  {M.}~\bibnamefont {Fortunato}}, \bibinfo {author} {\bibfnamefont
  {D.}~\bibnamefont {Pfau}}, \bibinfo {author} {\bibfnamefont {L.~R.}\
  \bibnamefont {Castellanos}}, \bibinfo {author} {\bibfnamefont
  {S.}~\bibnamefont {Petersen}}, \bibinfo {author} {\bibfnamefont {A.~W.~R.}\
  \bibnamefont {Nelson}}, \bibinfo {author} {\bibfnamefont {P.}~\bibnamefont
  {Kohli}}, \bibinfo {author} {\bibfnamefont {P.}~\bibnamefont
  {Mori-Sánchez}}, \bibinfo {author} {\bibfnamefont {D.}~\bibnamefont
  {Hassabis}}, \ and\ \bibinfo {author} {\bibfnamefont {A.~J.}\ \bibnamefont
  {Cohen}},\ }\bibfield  {title} {\enquote {\bibinfo {title} {Pushing the
  frontiers of density functionals by solving the fractional electron
  problem},}\ }\href {\doibase 10.1126/science.abj6511} {\bibfield  {journal}
  {\bibinfo  {journal} {Science}\ }\textbf {\bibinfo {volume} {374}},\ \bibinfo
  {pages} {1385--1389} (\bibinfo {year} {2021})}\BibitemShut {NoStop}%
\bibitem [{\citenamefont {Li}\ \emph {et~al.}(2021)\citenamefont {Li},
  \citenamefont {Hoyer}, \citenamefont {Pederson}, \citenamefont {Sun},
  \citenamefont {Cubuk}, \citenamefont {Riley},\ and\ \citenamefont
  {Burke}}]{Li2021}%
  \BibitemOpen
  \bibfield  {author} {\bibinfo {author} {\bibfnamefont {L.}~\bibnamefont
  {Li}}, \bibinfo {author} {\bibfnamefont {S.}~\bibnamefont {Hoyer}}, \bibinfo
  {author} {\bibfnamefont {R.}~\bibnamefont {Pederson}}, \bibinfo {author}
  {\bibfnamefont {R.}~\bibnamefont {Sun}}, \bibinfo {author} {\bibfnamefont
  {E.~D.}\ \bibnamefont {Cubuk}}, \bibinfo {author} {\bibfnamefont
  {P.}~\bibnamefont {Riley}}, \ and\ \bibinfo {author} {\bibfnamefont
  {K.}~\bibnamefont {Burke}},\ }\bibfield  {title} {\enquote {\bibinfo {title}
  {Kohn-sham equations as regularizer: Building prior knowledge into
  machine-learned physics},}\ }\href
  {https://link.aps.org/doi/10.1103/PhysRevLett.126.036401} {\bibfield
  {journal} {\bibinfo  {journal} {Phys. Rev. Lett.}\ }\textbf {\bibinfo
  {volume} {126}},\ \bibinfo {pages} {036401} (\bibinfo {year}
  {2021})}\BibitemShut {NoStop}%
\bibitem [{\citenamefont {Kasim}\ and\ \citenamefont
  {Vinko}(2021)}]{Kasim2021}%
  \BibitemOpen
  \bibfield  {author} {\bibinfo {author} {\bibfnamefont {M.~F.}\ \bibnamefont
  {Kasim}}\ and\ \bibinfo {author} {\bibfnamefont {S.~M.}\ \bibnamefont
  {Vinko}},\ }\bibfield  {title} {\enquote {\bibinfo {title} {Learning the
  exchange-correlation functional from nature with fully differentiable density
  functional theory},}\ }\href
  {https://link.aps.org/doi/10.1103/PhysRevLett.127.126403} {\bibfield
  {journal} {\bibinfo  {journal} {Phys. Rev. Lett.}\ }\textbf {\bibinfo
  {volume} {127}},\ \bibinfo {pages} {126403} (\bibinfo {year}
  {2021})}\BibitemShut {NoStop}%
\bibitem [{\citenamefont {Sun}\ \emph {et~al.}(2018)\citenamefont {Sun},
  \citenamefont {Berkelbach}, \citenamefont {Blunt}, \citenamefont {Booth},
  \citenamefont {Guo}, \citenamefont {Li}, \citenamefont {Liu}, \citenamefont
  {McClain}, \citenamefont {Sayfutyarova}, \citenamefont {Sharma},
  \citenamefont {Wouters},\ and\ \citenamefont {Chan}}]{Sun2018}%
  \BibitemOpen
  \bibfield  {author} {\bibinfo {author} {\bibfnamefont {Q.}~\bibnamefont
  {Sun}}, \bibinfo {author} {\bibfnamefont {T.~C.}\ \bibnamefont {Berkelbach}},
  \bibinfo {author} {\bibfnamefont {N.~S.}\ \bibnamefont {Blunt}}, \bibinfo
  {author} {\bibfnamefont {G.~H.}\ \bibnamefont {Booth}}, \bibinfo {author}
  {\bibfnamefont {S.}~\bibnamefont {Guo}}, \bibinfo {author} {\bibfnamefont
  {Z.}~\bibnamefont {Li}}, \bibinfo {author} {\bibfnamefont {J.}~\bibnamefont
  {Liu}}, \bibinfo {author} {\bibfnamefont {J.~D.}\ \bibnamefont {McClain}},
  \bibinfo {author} {\bibfnamefont {E.~R.}\ \bibnamefont {Sayfutyarova}},
  \bibinfo {author} {\bibfnamefont {S.}~\bibnamefont {Sharma}}, \bibinfo
  {author} {\bibfnamefont {S.}~\bibnamefont {Wouters}}, \ and\ \bibinfo
  {author} {\bibfnamefont {G.~K.-L.}\ \bibnamefont {Chan}},\ }\bibfield
  {title} {\enquote {\bibinfo {title} {Pyscf: the python-based simulations of
  chemistry framework},}\ }\href {https://doi.org/10.1002/wcms.1340} {\bibfield
   {journal} {\bibinfo  {journal} {WIREs Comput Mol Sci}\ }\textbf {\bibinfo
  {volume} {8}},\ \bibinfo {pages} {e1340} (\bibinfo {year}
  {2018})}\BibitemShut {NoStop}%
\bibitem [{\citenamefont {Sun}(2015)}]{Sun2015}%
  \BibitemOpen
  \bibfield  {author} {\bibinfo {author} {\bibfnamefont {Q.}~\bibnamefont
  {Sun}},\ }\bibfield  {title} {\enquote {\bibinfo {title} {Libcint: An
  efficient general integral library for gaussian basis functions},}\ }\href
  {https://doi.org/10.1002/jcc.23981} {\bibfield  {journal} {\bibinfo
  {journal} {J. Comput. Chem.}\ }\textbf {\bibinfo {volume} {36}},\ \bibinfo
  {pages} {1664--1671} (\bibinfo {year} {2015})}\BibitemShut {NoStop}%
\bibitem [{\citenamefont {Li}\ and\ \citenamefont
  {Chan}(2025{\natexlab{a}})}]{Li2025}%
  \BibitemOpen
  \bibfield  {author} {\bibinfo {author} {\bibfnamefont {C.}~\bibnamefont
  {Li}}\ and\ \bibinfo {author} {\bibfnamefont {G.~K.-L.}\ \bibnamefont
  {Chan}},\ }\bibfield  {title} {\enquote {\bibinfo {title} {Accurate qm/mm
  molecular dynamics for periodic systems in gpu4pyscf with applications to
  enzyme catalysis},}\ }\href {\doibase 10.1021/acs.jctc.4c01698} {\bibfield
  {journal} {\bibinfo  {journal} {J. Chem. Theory Comput.}\ }\textbf {\bibinfo
  {volume} {21}},\ \bibinfo {pages} {803--816} (\bibinfo {year}
  {2025}{\natexlab{a}})}\BibitemShut {NoStop}%
\bibitem [{\citenamefont {King}\ and\ \citenamefont
  {Gagliardi}(2021)}]{King2021}%
  \BibitemOpen
  \bibfield  {author} {\bibinfo {author} {\bibfnamefont {D.~S.}\ \bibnamefont
  {King}}\ and\ \bibinfo {author} {\bibfnamefont {L.}~\bibnamefont
  {Gagliardi}},\ }\bibfield  {title} {\enquote {\bibinfo {title} {A
  ranked-orbital approach to select active spaces for high-throughput
  multireference computation},}\ }\href {\doibase 10.1021/acs.jctc.1c00037}
  {\bibfield  {journal} {\bibinfo  {journal} {J. Chem. Theory Comput.}\
  }\textbf {\bibinfo {volume} {17}},\ \bibinfo {pages} {2817--2831} (\bibinfo
  {year} {2021})}\BibitemShut {NoStop}%
\bibitem [{\citenamefont {Golub}\ \emph {et~al.}(2021)\citenamefont {Golub},
  \citenamefont {Antalik}, \citenamefont {Veis},\ and\ \citenamefont
  {Brabec}}]{Golub2021}%
  \BibitemOpen
  \bibfield  {author} {\bibinfo {author} {\bibfnamefont {P.}~\bibnamefont
  {Golub}}, \bibinfo {author} {\bibfnamefont {A.}~\bibnamefont {Antalik}},
  \bibinfo {author} {\bibfnamefont {L.}~\bibnamefont {Veis}}, \ and\ \bibinfo
  {author} {\bibfnamefont {J.}~\bibnamefont {Brabec}},\ }\bibfield  {title}
  {\enquote {\bibinfo {title} {Machine learning-assisted selection of active
  spaces for strongly correlated transition metal systems},}\ }\href {\doibase
  10.1021/acs.jctc.1c00235} {\bibfield  {journal} {\bibinfo  {journal} {J.
  Chem. Theory Comput.}\ }\textbf {\bibinfo {volume} {17}},\ \bibinfo {pages}
  {6053--6072} (\bibinfo {year} {2021})}\BibitemShut {NoStop}%
\bibitem [{\citenamefont {Lei}, \citenamefont {Suo},\ and\ \citenamefont
  {Liu}(2021)}]{Lei2021}%
  \BibitemOpen
  \bibfield  {author} {\bibinfo {author} {\bibfnamefont {Y.}~\bibnamefont
  {Lei}}, \bibinfo {author} {\bibfnamefont {B.}~\bibnamefont {Suo}}, \ and\
  \bibinfo {author} {\bibfnamefont {W.}~\bibnamefont {Liu}},\ }\bibfield
  {title} {\enquote {\bibinfo {title} {icas: Imposed automatic selection and
  localization of complete active spaces},}\ }\href {\doibase
  10.1021/acs.jctc.1c00456} {\bibfield  {journal} {\bibinfo  {journal} {J.
  Chem. Theory Comput.}\ }\textbf {\bibinfo {volume} {17}},\ \bibinfo {pages}
  {4846--4859} (\bibinfo {year} {2021})}\BibitemShut {NoStop}%
\bibitem [{\citenamefont {Kolodzeiski}\ and\ \citenamefont
  {Stein}(2023)}]{Kolodzeiski2023}%
  \BibitemOpen
  \bibfield  {author} {\bibinfo {author} {\bibfnamefont {E.}~\bibnamefont
  {Kolodzeiski}}\ and\ \bibinfo {author} {\bibfnamefont {C.~J.}\ \bibnamefont
  {Stein}},\ }\bibfield  {title} {\enquote {\bibinfo {title} {Automated,
  consistent, and even-handed selection of active orbital spaces for quantum
  embedding},}\ }\href {\doibase 10.1021/acs.jctc.3c00653} {\bibfield
  {journal} {\bibinfo  {journal} {J. Chem. Theory Comput.}\ }\textbf {\bibinfo
  {volume} {19}},\ \bibinfo {pages} {6643--6655} (\bibinfo {year}
  {2023})}\BibitemShut {NoStop}%
\bibitem [{\citenamefont {Sayfutyarova}\ \emph {et~al.}(2017)\citenamefont
  {Sayfutyarova}, \citenamefont {Sun}, \citenamefont {Chan},\ and\
  \citenamefont {Knizia}}]{Sayfutyarova2017}%
  \BibitemOpen
  \bibfield  {author} {\bibinfo {author} {\bibfnamefont {E.~R.}\ \bibnamefont
  {Sayfutyarova}}, \bibinfo {author} {\bibfnamefont {Q.}~\bibnamefont {Sun}},
  \bibinfo {author} {\bibfnamefont {G.~K.-L.}\ \bibnamefont {Chan}}, \ and\
  \bibinfo {author} {\bibfnamefont {G.}~\bibnamefont {Knizia}},\ }\bibfield
  {title} {\enquote {\bibinfo {title} {Automated construction of molecular
  active spaces from atomic valence orbitals},}\ }\href {\doibase
  10.1021/acs.jctc.7b00128} {\bibfield  {journal} {\bibinfo  {journal} {J.
  Chem. Theory Comput.}\ }\textbf {\bibinfo {volume} {13}},\ \bibinfo {pages}
  {4063--4078} (\bibinfo {year} {2017})}\BibitemShut {NoStop}%
\bibitem [{\citenamefont {Zou}\ \emph {et~al.}(2020)\citenamefont {Zou},
  \citenamefont {Niu}, \citenamefont {Ma}, \citenamefont {Li},\ and\
  \citenamefont {Fang}}]{Zou2020}%
  \BibitemOpen
  \bibfield  {author} {\bibinfo {author} {\bibfnamefont {J.}~\bibnamefont
  {Zou}}, \bibinfo {author} {\bibfnamefont {K.}~\bibnamefont {Niu}}, \bibinfo
  {author} {\bibfnamefont {H.}~\bibnamefont {Ma}}, \bibinfo {author}
  {\bibfnamefont {S.}~\bibnamefont {Li}}, \ and\ \bibinfo {author}
  {\bibfnamefont {W.}~\bibnamefont {Fang}},\ }\bibfield  {title} {\enquote
  {\bibinfo {title} {Automatic selection of active orbitals from generalized
  valence bond orbitals},}\ }\href {\doibase 10.1021/acs.jpca.0c05216}
  {\bibfield  {journal} {\bibinfo  {journal} {J. Phys. Chem. A}\ }\textbf
  {\bibinfo {volume} {124}},\ \bibinfo {pages} {8321--8329} (\bibinfo {year}
  {2020})}\BibitemShut {NoStop}%
\bibitem [{\citenamefont {Bao}\ and\ \citenamefont {Truhlar}(2019)}]{Bao2019}%
  \BibitemOpen
  \bibfield  {author} {\bibinfo {author} {\bibfnamefont {J.~J.}\ \bibnamefont
  {Bao}}\ and\ \bibinfo {author} {\bibfnamefont {D.~G.}\ \bibnamefont
  {Truhlar}},\ }\bibfield  {title} {\enquote {\bibinfo {title} {Automatic
  active space selection for calculating electronic excitation energies based
  on high-spin unrestricted hartree-fock orbitals},}\ }\href {\doibase
  10.1021/acs.jctc.9b00535} {\bibfield  {journal} {\bibinfo  {journal} {J.
  Chem. Theory Comput.}\ }\textbf {\bibinfo {volume} {15}},\ \bibinfo {pages}
  {5308--5318} (\bibinfo {year} {2019})}\BibitemShut {NoStop}%
\bibitem [{\citenamefont {Sun}(2017)}]{Sun2017}%
  \BibitemOpen
  \bibfield  {author} {\bibinfo {author} {\bibfnamefont {Q.}~\bibnamefont
  {Sun}},\ }\href {https://arxiv.org/abs/1610.08423} {\enquote {\bibinfo
  {title} {Co-iterative augmented hessian method for orbital optimization},}\ }
  (\bibinfo {year} {2017}),\ \Eprint {http://arxiv.org/abs/1610.08423}
  {arXiv:1610.08423 [physics.chem-ph]} \BibitemShut {NoStop}%
\bibitem [{\citenamefont {Neese}(2000)}]{Neese2000}%
  \BibitemOpen
  \bibfield  {author} {\bibinfo {author} {\bibfnamefont {F.}~\bibnamefont
  {Neese}},\ }\bibfield  {title} {\enquote {\bibinfo {title} {Approximate
  second-order scf convergence for spin unrestricted wavefunctions},}\ }\href
  {https://www.sciencedirect.com/science/article/pii/S000926140000662X}
  {\bibfield  {journal} {\bibinfo  {journal} {Chemical Physics Letters}\
  }\textbf {\bibinfo {volume} {325}},\ \bibinfo {pages} {93--98} (\bibinfo
  {year} {2000})}\BibitemShut {NoStop}%
\bibitem [{\citenamefont {Thøgersen}\ \emph {et~al.}(2005)\citenamefont
  {Thøgersen}, \citenamefont {Olsen}, \citenamefont {Köhn}, \citenamefont
  {Jørgensen}, \citenamefont {Sałek},\ and\ \citenamefont
  {Helgaker}}]{Thoegersen2005}%
  \BibitemOpen
  \bibfield  {author} {\bibinfo {author} {\bibfnamefont {L.}~\bibnamefont
  {Thøgersen}}, \bibinfo {author} {\bibfnamefont {J.}~\bibnamefont {Olsen}},
  \bibinfo {author} {\bibfnamefont {A.}~\bibnamefont {Köhn}}, \bibinfo
  {author} {\bibfnamefont {P.}~\bibnamefont {Jørgensen}}, \bibinfo {author}
  {\bibfnamefont {P.}~\bibnamefont {Sałek}}, \ and\ \bibinfo {author}
  {\bibfnamefont {T.}~\bibnamefont {Helgaker}},\ }\bibfield  {title} {\enquote
  {\bibinfo {title} {The trust-region self-consistent field method in kohn-sham
  density-functional theory},}\ }\href {https://doi.org/10.1063/1.1989311}
  {\bibfield  {journal} {\bibinfo  {journal} {J. Chem. Phys.}\ }\textbf
  {\bibinfo {volume} {123}},\ \bibinfo {pages} {074103} (\bibinfo {year}
  {2005})}\BibitemShut {NoStop}%
\bibitem [{\citenamefont {Fischer}\ and\ \citenamefont
  {Almlof}(1992)}]{Fischer1992}%
  \BibitemOpen
  \bibfield  {author} {\bibinfo {author} {\bibfnamefont {T.~H.}\ \bibnamefont
  {Fischer}}\ and\ \bibinfo {author} {\bibfnamefont {J.}~\bibnamefont
  {Almlof}},\ }\bibfield  {title} {\enquote {\bibinfo {title} {General methods
  for geometry and wave function optimization},}\ }\href {\doibase
  10.1021/j100203a036} {\bibfield  {journal} {\bibinfo  {journal} {J. Phys.
  Chem.}\ }\textbf {\bibinfo {volume} {96}},\ \bibinfo {pages} {9768--9774}
  (\bibinfo {year} {1992})}\BibitemShut {NoStop}%
\bibitem [{\citenamefont {Slattery}\ \emph {et~al.}(2024)\citenamefont
  {Slattery}, \citenamefont {Surjuse}, \citenamefont {Peterson}, \citenamefont
  {Penchoff},\ and\ \citenamefont {Valeev}}]{Slattery2024}%
  \BibitemOpen
  \bibfield  {author} {\bibinfo {author} {\bibfnamefont {S.~A.}\ \bibnamefont
  {Slattery}}, \bibinfo {author} {\bibfnamefont {K.~A.}\ \bibnamefont
  {Surjuse}}, \bibinfo {author} {\bibfnamefont {C.~C.}\ \bibnamefont
  {Peterson}}, \bibinfo {author} {\bibfnamefont {D.~A.}\ \bibnamefont
  {Penchoff}}, \ and\ \bibinfo {author} {\bibfnamefont {E.~F.}\ \bibnamefont
  {Valeev}},\ }\bibfield  {title} {\enquote {\bibinfo {title} {Economical
  quasi-newton unitary optimization of electronic orbitals},}\ }\href
  {http://dx.doi.org/10.1039/D3CP05557D} {\bibfield  {journal} {\bibinfo
  {journal} {Phys. Chem. Chem. Phys.}\ }\textbf {\bibinfo {volume} {26}},\
  \bibinfo {pages} {6557--6573} (\bibinfo {year} {2024})}\BibitemShut {NoStop}%
\bibitem [{\citenamefont {Weigend}(2008)}]{Weigend2008}%
  \BibitemOpen
  \bibfield  {author} {\bibinfo {author} {\bibfnamefont {F.}~\bibnamefont
  {Weigend}},\ }\bibfield  {title} {\enquote {\bibinfo {title} {Hartree-fock
  exchange fitting basis sets for h to rn},}\ }\href
  {https://doi.org/10.1002/jcc.20702} {\bibfield  {journal} {\bibinfo
  {journal} {J. Comput. Chem.}\ }\textbf {\bibinfo {volume} {29}},\ \bibinfo
  {pages} {167--175} (\bibinfo {year} {2008})}\BibitemShut {NoStop}%
\bibitem [{\citenamefont {Kühne}\ \emph {et~al.}(2020)\citenamefont {Kühne},
  \citenamefont {Iannuzzi}, \citenamefont {Del~Ben}, \citenamefont {Rybkin},
  \citenamefont {Seewald}, \citenamefont {Stein}, \citenamefont {Laino},
  \citenamefont {Khaliullin}, \citenamefont {Schütt}, \citenamefont
  {Schiffmann}, \citenamefont {Golze}, \citenamefont {Wilhelm}, \citenamefont
  {Chulkov}, \citenamefont {Bani-Hashemian}, \citenamefont {Weber},
  \citenamefont {Borštnik}, \citenamefont {Taillefumier}, \citenamefont
  {Jakobovits}, \citenamefont {Lazzaro}, \citenamefont {Pabst}, \citenamefont
  {Müller}, \citenamefont {Schade}, \citenamefont {Guidon}, \citenamefont
  {Andermatt}, \citenamefont {Holmberg}, \citenamefont {Schenter},
  \citenamefont {Hehn}, \citenamefont {Bussy}, \citenamefont {Belleflamme},
  \citenamefont {Tabacchi}, \citenamefont {Glöß}, \citenamefont {Lass},
  \citenamefont {Bethune}, \citenamefont {Mundy}, \citenamefont {Plessl},
  \citenamefont {Watkins}, \citenamefont {VandeVondele}, \citenamefont
  {Krack},\ and\ \citenamefont {Hutter}}]{Kuehne2020}%
  \BibitemOpen
  \bibfield  {author} {\bibinfo {author} {\bibfnamefont {T.~D.}\ \bibnamefont
  {Kühne}}, \bibinfo {author} {\bibfnamefont {M.}~\bibnamefont {Iannuzzi}},
  \bibinfo {author} {\bibfnamefont {M.}~\bibnamefont {Del~Ben}}, \bibinfo
  {author} {\bibfnamefont {V.~V.}\ \bibnamefont {Rybkin}}, \bibinfo {author}
  {\bibfnamefont {P.}~\bibnamefont {Seewald}}, \bibinfo {author} {\bibfnamefont
  {F.}~\bibnamefont {Stein}}, \bibinfo {author} {\bibfnamefont
  {T.}~\bibnamefont {Laino}}, \bibinfo {author} {\bibfnamefont {R.~Z.}\
  \bibnamefont {Khaliullin}}, \bibinfo {author} {\bibfnamefont
  {O.}~\bibnamefont {Schütt}}, \bibinfo {author} {\bibfnamefont
  {F.}~\bibnamefont {Schiffmann}}, \bibinfo {author} {\bibfnamefont
  {D.}~\bibnamefont {Golze}}, \bibinfo {author} {\bibfnamefont
  {J.}~\bibnamefont {Wilhelm}}, \bibinfo {author} {\bibfnamefont
  {S.}~\bibnamefont {Chulkov}}, \bibinfo {author} {\bibfnamefont {M.~H.}\
  \bibnamefont {Bani-Hashemian}}, \bibinfo {author} {\bibfnamefont
  {V.}~\bibnamefont {Weber}}, \bibinfo {author} {\bibfnamefont
  {U.}~\bibnamefont {Borštnik}}, \bibinfo {author} {\bibfnamefont
  {M.}~\bibnamefont {Taillefumier}}, \bibinfo {author} {\bibfnamefont {A.~S.}\
  \bibnamefont {Jakobovits}}, \bibinfo {author} {\bibfnamefont
  {A.}~\bibnamefont {Lazzaro}}, \bibinfo {author} {\bibfnamefont
  {H.}~\bibnamefont {Pabst}}, \bibinfo {author} {\bibfnamefont
  {T.}~\bibnamefont {Müller}}, \bibinfo {author} {\bibfnamefont
  {R.}~\bibnamefont {Schade}}, \bibinfo {author} {\bibfnamefont
  {M.}~\bibnamefont {Guidon}}, \bibinfo {author} {\bibfnamefont
  {S.}~\bibnamefont {Andermatt}}, \bibinfo {author} {\bibfnamefont
  {N.}~\bibnamefont {Holmberg}}, \bibinfo {author} {\bibfnamefont {G.~K.}\
  \bibnamefont {Schenter}}, \bibinfo {author} {\bibfnamefont {A.}~\bibnamefont
  {Hehn}}, \bibinfo {author} {\bibfnamefont {A.}~\bibnamefont {Bussy}},
  \bibinfo {author} {\bibfnamefont {F.}~\bibnamefont {Belleflamme}}, \bibinfo
  {author} {\bibfnamefont {G.}~\bibnamefont {Tabacchi}}, \bibinfo {author}
  {\bibfnamefont {A.}~\bibnamefont {Glöß}}, \bibinfo {author} {\bibfnamefont
  {M.}~\bibnamefont {Lass}}, \bibinfo {author} {\bibfnamefont {I.}~\bibnamefont
  {Bethune}}, \bibinfo {author} {\bibfnamefont {C.~J.}\ \bibnamefont {Mundy}},
  \bibinfo {author} {\bibfnamefont {C.}~\bibnamefont {Plessl}}, \bibinfo
  {author} {\bibfnamefont {M.}~\bibnamefont {Watkins}}, \bibinfo {author}
  {\bibfnamefont {J.}~\bibnamefont {VandeVondele}}, \bibinfo {author}
  {\bibfnamefont {M.}~\bibnamefont {Krack}}, \ and\ \bibinfo {author}
  {\bibfnamefont {J.}~\bibnamefont {Hutter}},\ }\bibfield  {title} {\enquote
  {\bibinfo {title} {Cp2k: An electronic structure and molecular dynamics
  software package - quickstep: Efficient and accurate electronic structure
  calculations},}\ }\href {https://doi.org/10.1063/5.0007045} {\bibfield
  {journal} {\bibinfo  {journal} {J. Chem. Phys.}\ }\textbf {\bibinfo {volume}
  {152}},\ \bibinfo {pages} {194103} (\bibinfo {year} {2020})}\BibitemShut
  {NoStop}%
\bibitem [{\citenamefont {Li}\ and\ \citenamefont
  {Chan}(2025{\natexlab{b}})}]{Li2025b}%
  \BibitemOpen
  \bibfield  {author} {\bibinfo {author} {\bibfnamefont {R.}~\bibnamefont
  {Li}}\ and\ \bibinfo {author} {\bibfnamefont {G.}~\bibnamefont {Chan}},\
  }\href@noop {} {\enquote {\bibinfo {title} {Manuscript in preparation on gpu
  acceleration for multigrid integration algorithm},}\ } (\bibinfo {year}
  {2025}{\natexlab{b}}),\ \bibinfo {note} {manuscript in
  preparation}\BibitemShut {NoStop}%
\bibitem [{\citenamefont {Zhang}\ and\ \citenamefont {Chan}(2025)}]{Zhang2025}%
  \BibitemOpen
  \bibfield  {author} {\bibinfo {author} {\bibfnamefont {X.}~\bibnamefont
  {Zhang}}\ and\ \bibinfo {author} {\bibfnamefont {G.}~\bibnamefont {Chan}},\
  }\href@noop {} {\enquote {\bibinfo {title} {Manuscript in preparation on
  multigrid integration algorithm in pyscf},}\ } (\bibinfo {year} {2025}),\
  \bibinfo {note} {manuscript in preparation}\BibitemShut {NoStop}%
\bibitem [{\citenamefont {Vahtras}, \citenamefont {Almlöf},\ and\
  \citenamefont {Feyereisen}(1993)}]{Vahtras1993}%
  \BibitemOpen
  \bibfield  {author} {\bibinfo {author} {\bibfnamefont {O.}~\bibnamefont
  {Vahtras}}, \bibinfo {author} {\bibfnamefont {J.}~\bibnamefont {Almlöf}}, \
  and\ \bibinfo {author} {\bibfnamefont {M.~W.}\ \bibnamefont {Feyereisen}},\
  }\bibfield  {title} {\enquote {\bibinfo {title} {Integral approximations for
  lcao-scf calculations},}\ }\href
  {https://www.sciencedirect.com/science/article/pii/0009261493891517}
  {\bibfield  {journal} {\bibinfo  {journal} {Chemical Physics Letters}\
  }\textbf {\bibinfo {volume} {213}},\ \bibinfo {pages} {514--518} (\bibinfo
  {year} {1993})}\BibitemShut {NoStop}%
\bibitem [{\citenamefont {Hill}(2013)}]{Hill2013}%
  \BibitemOpen
  \bibfield  {author} {\bibinfo {author} {\bibfnamefont {J.~G.}\ \bibnamefont
  {Hill}},\ }\bibfield  {title} {\enquote {\bibinfo {title} {Gaussian basis
  sets for molecular applications},}\ }\href
  {https://doi.org/10.1002/qua.24355} {\bibfield  {journal} {\bibinfo
  {journal} {Int. J. Quantum Chem.}\ }\textbf {\bibinfo {volume} {113}},\
  \bibinfo {pages} {21--34} (\bibinfo {year} {2013})}\BibitemShut {NoStop}%
\bibitem [{\citenamefont {Eichkorn}\ \emph {et~al.}(1995)\citenamefont
  {Eichkorn}, \citenamefont {Treutler}, \citenamefont {Öhm}, \citenamefont
  {Häser},\ and\ \citenamefont {Ahlrichs}}]{Eichkorn1995}%
  \BibitemOpen
  \bibfield  {author} {\bibinfo {author} {\bibfnamefont {K.}~\bibnamefont
  {Eichkorn}}, \bibinfo {author} {\bibfnamefont {O.}~\bibnamefont {Treutler}},
  \bibinfo {author} {\bibfnamefont {H.}~\bibnamefont {Öhm}}, \bibinfo {author}
  {\bibfnamefont {M.}~\bibnamefont {Häser}}, \ and\ \bibinfo {author}
  {\bibfnamefont {R.}~\bibnamefont {Ahlrichs}},\ }\bibfield  {title} {\enquote
  {\bibinfo {title} {Auxiliary basis sets to approximate coulomb potentials},}\
  }\href {https://www.sciencedirect.com/science/article/pii/000926149500621A}
  {\bibfield  {journal} {\bibinfo  {journal} {Chemical Physics Letters}\
  }\textbf {\bibinfo {volume} {240}},\ \bibinfo {pages} {283--290} (\bibinfo
  {year} {1995})}\BibitemShut {NoStop}%
\bibitem [{\citenamefont {Eichkorn}\ \emph {et~al.}(1997)\citenamefont
  {Eichkorn}, \citenamefont {Weigend}, \citenamefont {Treutler},\ and\
  \citenamefont {Ahlrichs}}]{Eichkorn1997}%
  \BibitemOpen
  \bibfield  {author} {\bibinfo {author} {\bibfnamefont {K.}~\bibnamefont
  {Eichkorn}}, \bibinfo {author} {\bibfnamefont {F.}~\bibnamefont {Weigend}},
  \bibinfo {author} {\bibfnamefont {O.}~\bibnamefont {Treutler}}, \ and\
  \bibinfo {author} {\bibfnamefont {R.}~\bibnamefont {Ahlrichs}},\ }\bibfield
  {title} {\enquote {\bibinfo {title} {Auxiliary basis sets for main row atoms
  and transition metals and their use to approximate coulomb potentials},}\
  }\href {https://doi.org/10.1007/s002140050244} {\bibfield  {journal}
  {\bibinfo  {journal} {Theoretical Chemistry Accounts}\ }\textbf {\bibinfo
  {volume} {97}},\ \bibinfo {pages} {119--124} (\bibinfo {year}
  {1997})}\BibitemShut {NoStop}%
\bibitem [{\citenamefont {Aquilante}\ \emph {et~al.}(2009)\citenamefont
  {Aquilante}, \citenamefont {Gagliardi}, \citenamefont {Pedersen},\ and\
  \citenamefont {Lindh}}]{Aquilante2009}%
  \BibitemOpen
  \bibfield  {author} {\bibinfo {author} {\bibfnamefont {F.}~\bibnamefont
  {Aquilante}}, \bibinfo {author} {\bibfnamefont {L.}~\bibnamefont
  {Gagliardi}}, \bibinfo {author} {\bibfnamefont {T.~B.}\ \bibnamefont
  {Pedersen}}, \ and\ \bibinfo {author} {\bibfnamefont {R.}~\bibnamefont
  {Lindh}},\ }\bibfield  {title} {\enquote {\bibinfo {title} {Atomic cholesky
  decompositions: A route to unbiased auxiliary basis sets for density fitting
  approximation with tunable accuracy and efficiency},}\ }\href
  {https://doi.org/10.1063/1.3116784} {\bibfield  {journal} {\bibinfo
  {journal} {J. Chem. Phys.}\ }\textbf {\bibinfo {volume} {130}},\ \bibinfo
  {pages} {154107} (\bibinfo {year} {2009})}\BibitemShut {NoStop}%
\bibitem [{\citenamefont {Lehtola}(2021)}]{Lehtola2021}%
  \BibitemOpen
  \bibfield  {author} {\bibinfo {author} {\bibfnamefont {S.}~\bibnamefont
  {Lehtola}},\ }\bibfield  {title} {\enquote {\bibinfo {title} {Straightforward
  and accurate automatic auxiliary basis set generation for molecular
  calculations with atomic orbital basis sets},}\ }\href {\doibase
  10.1021/acs.jctc.1c00607} {\bibfield  {journal} {\bibinfo  {journal} {J.
  Chem. Theory Comput.}\ }\textbf {\bibinfo {volume} {17}},\ \bibinfo {pages}
  {6886--6900} (\bibinfo {year} {2021})}\BibitemShut {NoStop}%
\bibitem [{\citenamefont {Lehtola}(2023)}]{Lehtola2023}%
  \BibitemOpen
  \bibfield  {author} {\bibinfo {author} {\bibfnamefont {S.}~\bibnamefont
  {Lehtola}},\ }\bibfield  {title} {\enquote {\bibinfo {title} {Automatic
  generation of accurate and cost-efficient auxiliary basis sets},}\ }\href
  {\doibase 10.1021/acs.jctc.3c00670} {\bibfield  {journal} {\bibinfo
  {journal} {J. Chem. Theory Comput.}\ }\textbf {\bibinfo {volume} {19}},\
  \bibinfo {pages} {6242--6254} (\bibinfo {year} {2023})}\BibitemShut {NoStop}%
\bibitem [{\citenamefont {Stoychev}, \citenamefont {Auer},\ and\ \citenamefont
  {Neese}(2017)}]{Stoychev2017}%
  \BibitemOpen
  \bibfield  {author} {\bibinfo {author} {\bibfnamefont {G.~L.}\ \bibnamefont
  {Stoychev}}, \bibinfo {author} {\bibfnamefont {A.~A.}\ \bibnamefont {Auer}},
  \ and\ \bibinfo {author} {\bibfnamefont {F.}~\bibnamefont {Neese}},\
  }\bibfield  {title} {\enquote {\bibinfo {title} {Automatic generation of
  auxiliary basis sets},}\ }\href {\doibase 10.1021/acs.jctc.6b01041}
  {\bibfield  {journal} {\bibinfo  {journal} {J. Chem. Theory Comput.}\
  }\textbf {\bibinfo {volume} {13}},\ \bibinfo {pages} {554--562} (\bibinfo
  {year} {2017})}\BibitemShut {NoStop}%
\bibitem [{\citenamefont {Aquilante}, \citenamefont {Lindh},\ and\
  \citenamefont {Bondo~Pedersen}(2007)}]{Aquilante2007}%
  \BibitemOpen
  \bibfield  {author} {\bibinfo {author} {\bibfnamefont {F.}~\bibnamefont
  {Aquilante}}, \bibinfo {author} {\bibfnamefont {R.}~\bibnamefont {Lindh}}, \
  and\ \bibinfo {author} {\bibfnamefont {T.}~\bibnamefont {Bondo~Pedersen}},\
  }\bibfield  {title} {\enquote {\bibinfo {title} {Unbiased auxiliary basis
  sets for accurate two-electron integral approximations},}\ }\href
  {https://doi.org/10.1063/1.2777146} {\bibfield  {journal} {\bibinfo
  {journal} {J. Chem. Phys.}\ }\textbf {\bibinfo {volume} {127}},\ \bibinfo
  {pages} {114107} (\bibinfo {year} {2007})}\BibitemShut {NoStop}%
\bibitem [{\citenamefont {Yang}, \citenamefont {Rendell},\ and\ \citenamefont
  {Frisch}(2007)}]{Yang2007}%
  \BibitemOpen
  \bibfield  {author} {\bibinfo {author} {\bibfnamefont {R.}~\bibnamefont
  {Yang}}, \bibinfo {author} {\bibfnamefont {A.~P.}\ \bibnamefont {Rendell}}, \
  and\ \bibinfo {author} {\bibfnamefont {M.~J.}\ \bibnamefont {Frisch}},\
  }\bibfield  {title} {\enquote {\bibinfo {title} {Automatically generated
  coulomb fitting basis sets: Design and accuracy for systems containing h to
  kr},}\ }\href {https://doi.org/10.1063/1.2752807} {\bibfield  {journal}
  {\bibinfo  {journal} {J. Chem. Phys.}\ }\textbf {\bibinfo {volume} {127}},\
  \bibinfo {pages} {074102} (\bibinfo {year} {2007})}\BibitemShut {NoStop}%
\bibitem [{\citenamefont {Díaz-Tinoco}\ \emph {et~al.}(2025)\citenamefont
  {Díaz-Tinoco}, \citenamefont {Flores-Moreno}, \citenamefont
  {Zúñiga-Gutiérrez},\ and\ \citenamefont {Köster}}]{Diaz-Tinoco2025}%
  \BibitemOpen
  \bibfield  {author} {\bibinfo {author} {\bibfnamefont {M.}~\bibnamefont
  {Díaz-Tinoco}}, \bibinfo {author} {\bibfnamefont {R.}~\bibnamefont
  {Flores-Moreno}}, \bibinfo {author} {\bibfnamefont {B.~A.}\ \bibnamefont
  {Zúñiga-Gutiérrez}}, \ and\ \bibinfo {author} {\bibfnamefont {A.~M.}\
  \bibnamefont {Köster}},\ }\bibfield  {title} {\enquote {\bibinfo {title}
  {Automatic generation of even-tempered auxiliary basis sets with shared
  exponents for density fitting},}\ }\href {\doibase 10.1021/acs.jctc.4c01555}
  {\bibfield  {journal} {\bibinfo  {journal} {J. Chem. Theory Comput.}\
  }\textbf {\bibinfo {volume} {21}},\ \bibinfo {pages} {2338--2352} (\bibinfo
  {year} {2025})}\BibitemShut {NoStop}%
\bibitem [{\citenamefont {Developers}(2025{\natexlab{a}})}]{psi42025basissets}%
  \BibitemOpen
  \bibfield  {author} {\bibinfo {author} {\bibfnamefont {P.}~\bibnamefont
  {Developers}},\ }\href@noop {} {\enquote {\bibinfo {title} {Psi4 manual:
  Basis set families},}\ }\bibinfo {howpublished}
  {\url{https://psicode.org/psi4manual/master/basissets_byfamily.html#apdx-basisfamily}}
  (\bibinfo {year} {2025}{\natexlab{a}}),\ \bibinfo {note} {accessed:
  2025-05-29}\BibitemShut {NoStop}%
\bibitem [{\citenamefont {Sun}(2023)}]{Sun2023}%
  \BibitemOpen
  \bibfield  {author} {\bibinfo {author} {\bibfnamefont {Q.}~\bibnamefont
  {Sun}},\ }\bibfield  {title} {\enquote {\bibinfo {title} {Efficient
  hartree-fock exchange algorithm with coulomb range separation and long-range
  density fitting},}\ }\href {https://doi.org/10.1063/5.0178266} {\bibfield
  {journal} {\bibinfo  {journal} {J. Chem. Phys.}\ }\textbf {\bibinfo {volume}
  {159}},\ \bibinfo {pages} {224101} (\bibinfo {year} {2023})}\BibitemShut
  {NoStop}%
\bibitem [{\citenamefont {Boldo}\ and\ \citenamefont
  {Muller}(2011)}]{Boldo2011}%
  \BibitemOpen
  \bibfield  {author} {\bibinfo {author} {\bibfnamefont {S.}~\bibnamefont
  {Boldo}}\ and\ \bibinfo {author} {\bibfnamefont {J.-M.}\ \bibnamefont
  {Muller}},\ }\bibfield  {title} {\enquote {\bibinfo {title} {Exact and
  approximated error of the fma},}\ }\href {\doibase 10.1109/TC.2010.139}
  {\bibfield  {journal} {\bibinfo  {journal} {IEEE Trans. Comput.}\ }\textbf
  {\bibinfo {volume} {60}},\ \bibinfo {pages} {157–164} (\bibinfo {year}
  {2011})}\BibitemShut {NoStop}%
\bibitem [{\citenamefont {Muller}\ \emph {et~al.}(2010)\citenamefont {Muller},
  \citenamefont {Brisebarre}, \citenamefont {de~Dinechin}, \citenamefont
  {Jeannerod}, \citenamefont {Lef{\`e}vre}, \citenamefont {Melquiond},
  \citenamefont {Revol}, \citenamefont {Stehl{\'e}},\ and\ \citenamefont
  {Torres}}]{Muller2010}%
  \BibitemOpen
  \bibfield  {author} {\bibinfo {author} {\bibfnamefont {J.-M.}\ \bibnamefont
  {Muller}}, \bibinfo {author} {\bibfnamefont {N.}~\bibnamefont {Brisebarre}},
  \bibinfo {author} {\bibfnamefont {F.}~\bibnamefont {de~Dinechin}}, \bibinfo
  {author} {\bibfnamefont {C.-P.}\ \bibnamefont {Jeannerod}}, \bibinfo {author}
  {\bibfnamefont {V.}~\bibnamefont {Lef{\`e}vre}}, \bibinfo {author}
  {\bibfnamefont {G.}~\bibnamefont {Melquiond}}, \bibinfo {author}
  {\bibfnamefont {N.}~\bibnamefont {Revol}}, \bibinfo {author} {\bibfnamefont
  {D.}~\bibnamefont {Stehl{\'e}}}, \ and\ \bibinfo {author} {\bibfnamefont
  {S.}~\bibnamefont {Torres}},\ }\enquote {\bibinfo {title} {The fused
  multiply-add instruction},}\ in\ \href {\doibase 10.1007/978-0-8176-4705-6_5}
  {\emph {\bibinfo {booktitle} {Handbook of Floating-Point Arithmetic}}}\
  (\bibinfo  {publisher} {Birkh{\"a}user Boston},\ \bibinfo {address}
  {Boston},\ \bibinfo {year} {2010})\ pp.\ \bibinfo {pages}
  {151--179}\BibitemShut {NoStop}%
\bibitem [{\citenamefont {Sun}(2025{\natexlab{a}})}]{Sun2025a}%
  \BibitemOpen
  \bibfield  {author} {\bibinfo {author} {\bibfnamefont {Q.}~\bibnamefont
  {Sun}},\ }\bibfield  {title} {\enquote {\bibinfo {title} {Chapter 9 - program
  performance optimization},}\ }in\ \href
  {https://www.sciencedirect.com/science/article/pii/B9780443238376000181}
  {\emph {\bibinfo {booktitle} {Theoretical and Computational Chemistry}}},\
  Vol.~\bibinfo {volume} {23}\ (\bibinfo  {publisher} {Elsevier},\ \bibinfo
  {year} {2025})\ pp.\ \bibinfo {pages} {305--394}\BibitemShut {NoStop}%
\bibitem [{\citenamefont {Developers}(2025{\natexlab{b}})}]{Jax2025}%
  \BibitemOpen
  \bibfield  {author} {\bibinfo {author} {\bibfnamefont {J.}~\bibnamefont
  {Developers}},\ }\href@noop {} {\enquote {\bibinfo {title} {Jax pytrees
  documentation},}\ }\bibinfo {howpublished}
  {\url{https://docs.jax.dev/en/latest/pytrees.html}} (\bibinfo {year}
  {2025}{\natexlab{b}}),\ \bibinfo {note} {accessed: 2025-05-29}\BibitemShut
  {NoStop}%
\bibitem [{\citenamefont {Sun}(2025{\natexlab{b}})}]{Sun2025}%
  \BibitemOpen
  \bibfield  {author} {\bibinfo {author} {\bibfnamefont {Q.}~\bibnamefont
  {Sun}},\ }\bibfield  {title} {\enquote {\bibinfo {title} {Chapter 16 -
  molecular properties},}\ }in\ \href
  {https://www.sciencedirect.com/science/article/pii/B9780443238376000260}
  {\emph {\bibinfo {booktitle} {Theoretical and Computational Chemistry}}},\
  Vol.~\bibinfo {volume} {23}\ (\bibinfo  {publisher} {Elsevier},\ \bibinfo
  {year} {2025})\ pp.\ \bibinfo {pages} {675--712}\BibitemShut {NoStop}%
\end{thebibliography}%
\end{document}